\documentclass[letterpaper,aps,prd,twocolumn,tightenlines,preprintnumbers,nofootinbib,showkeys,superscriptaddress]{revtex4-1}
\pdfoutput=1

\usepackage[colorlinks=true,linkcolor=blue,citecolor=blue,urlcolor=blue]{hyperref}
\usepackage{natbib}
\usepackage[T1]{fontenc}
\usepackage{graphicx}
\usepackage{subcaption}
\usepackage{empheq}
\usepackage{slashed}
\usepackage{amsmath,float}
\usepackage{amsfonts}
\usepackage{amssymb}
\usepackage{xcolor}
\usepackage{framed}
\usepackage[toc,page]{appendix}
\usepackage{braket}
\usepackage{caption}
\usepackage{color}
\usepackage[normalem]{ulem}

 \DeclareUnicodeCharacter{202F}{FIX ME!!!!}

\def\beq{\begin{equation}}
\def\eeq{\end{equation}}
\def\barr{\begin{eqnarray}}
\def\earr{\end{eqnarray}}

\begin{document}
\title{  Light Dark Matter Detection and Neutrino Floor: Role of Anomalous $(g-2)_{\mu}$ 
% in $U(1)_{L_{\mu}-L_{\tau}}$ Model
}

\author{Kuldeep Deka}
\email{kuldeepdeka.physics@gmail.com}
\affiliation{Department of Physics and Astrophysics, University of Delhi, Delhi 110 007, India}

\author{Soumya Sadhukhan}
\email{soumya.sadhukhan@rkmrc.in}
\affiliation{Ramakrishna Mission Residential College (Autonomous), Narendrapur, Kolkata 700103, India} 
%\affiliation{Department of Physics and Astrophysics, University of Delhi, Delhi 110 007, India}

\author{Manvinder Pal Singh}
\email{manvinderpal666@yahoo.com}
\affiliation{Department of Physics and Astrophysics, University of Delhi, Delhi 110 007, India}

\begin{abstract}
%{\color{green}In this work, we explore the connection of light dark matter (DM) direct detection with modification of $(g-2)_{\mu}$, motivated by its recent anomalous measurement with $\sim 4 \sigma$ deviation from the SM.}
In this work, we explore the impact of dark matter (DM) relic density and direct detection constraints on a GeV scale DM in the context of recent anomalous muon magnetic moment $(g-2)_{\mu}$ measurement; a $ 5.1 \sigma$ discrepancy with the SM. 
%{\color{green}The $U(1)_{L_{\mu}-L_{\tau}}$ scenario modifies $(g-2)_{\mu}$ readily explaining the anomaly, and that restricts the $Z^{\prime}$ mass in the range of $20-200$~MeV. }
In $U(1)_{L_{\mu}-L_{\tau}}$ scenario the additional $Z'$ boson modifies the $(g-2)_{\mu}$ value readily explaining the discrepancy, which restricts the $Z^{\prime}$ mass in the range of $20-200$~MeV.
%{ \color{green}The neutrino floor is a critical component of the DM direct detection process, measuring the neutrino background to the DM signal.(Do we need to write previous sentence in the abstract, it seems it breaks the flow)} 
Bounds imposed on the $Z^{\prime}$ mass along with the gauge coupling, limit possible enhancement of the neutrino floor in an $U(1)_{L_{\mu}-L_{\tau}}$ model. 
Neutrino floor is enhanced for a lighter $Z^{\prime}$ inside the $(g-2)_{\mu}$ allowed parameter space, whereas for a heavier $Z^{\prime}$, enhancement is less significant. 
%This neutrino floor modification is included in the DM analysis which churns out novel parameter regions in the DM sector.
%
The $(g-2)_\mu$ constraint for the GeV scale Fermionic DM makes s-channel resonant annihilation insignificant, placing emphasis on a t-channel reliance to create the observed DM relic. 
Although a t-channel annihilation aided by relatively large couplings can explain the measured relic density, it increases the direct detection cross-section of the GeV DM. 
Consequently, super-GeV (with mass $1-10$~GeV) DM almost gets ruled out except for a small parameter region with heavier $Z^{\prime}$, whereas sub-GeV (with mass $0.1-1$~GeV) DM detection possibility remains bright with more detection possibility for heavier $Z^{\prime}$.
%{\color{green} Direct detection constraints when aided by $(g-2)_{\mu}$, constrains the GeV DM more, compared to the DM indirect detection measurements.} 
In our analysis, we have discovered that direct detection constraints have a greater impact on the GeV DM compared to indirect detection measurements. 

\end{abstract}

\maketitle

\section{Introduction}
One remarkable aspect that motivates us to look beyond the Standard Model (BSM) signature is the absence of a dark matter candidate inside the SM. 
As found out by astrophysical observations like galactic rotational curves and gravitational lensing etc, the majority ($\sim 85\%$) of the matter element of the Universe is some inert non-luminous matter called dark matter (DM), interacting only through gravitational interaction. 
Various particle-DM candidates have been incorporated in different beyond the standard model (BSM) theories like Inert Higgs Doublet, Right-handed neutrino, and Super-symmetry, to name a few. 
But to date, no conclusive observational evidence of such a particle is found either in the LHC, specifically designed to probe the TeV scale physics, or in the DM direct and indirect detection experiments. 

Particle candidates for DM are well motivated by the WIMP (weakly interacting massive particle) miracle, where a DM mass is expected to be at the TeV scale with interaction strength typically around that of the electroweak theory, to have correct DM relic density. 
%{\color{green} The search for the DM particles is on through different DM direct detection experiments, albeit with a renewed vigor directed to find DM particles at a lower mass scale.}/
 The DM particles are searched through the measurement of nuclear recoil in DM direct detection experiments, cosmic particle flux arising from DM annihilations in indirect detection, and the missing energy signal in collider experiments. 
Due to the lack of evidence of a DM at the TeV scale, the search continues, albeit with a renewed focus on finding DM particles at a lower mass scale. 
The GeV scale DM ( 0.1-10 GeV) is particularly interesting from the prospect of their experimental probe. 
 In this scale, DM is too light to be significantly sensitive to the LHC searches, which in general has played a crucial role in constraining the WIMPs. 
Moreover, the cosmological constraints are also comparatively relaxed for a GeV-scale DM. 
Therefore, the light DM sector at the GeV scale is relatively less constrained compared both to an ultralight DM and the WIMPs. 
It is thus imperative to probe the GeV regime with innovative probes.

 The presence of anomalies in the low-energy experimental data can provide us with crucial information to figure out the exact nature of BSM physics at lower scales. 
 One such anomaly, which has been persistent for a while and has recently gained renewed interest is the anomalous measurement of the muon magnetic moment, $(g-2)_\mu$. 
The preferred observable that we work on is defined as $ a_{\mu} = \frac{g-2}{2}$.  
Recently $(g-2)_\mu$ measurement at the Fermilab experiment has reported a deviation from the SM prediction of $\Delta a_{\mu}$ as~\cite{Muong-2:2023cdq}

$$\Delta a_\mu^{\rm (FNAL + BNL)_{2023}} = 249(48) \times 10^{-11}$$

%$$ \Delta a_{\mu} = a_{\mu}^{\rm exp} -a_{\mu}^{\rm SM} = (251 \pm 59) \times 10^{-11}  $$,
which is equivalent to $5.1\sigma$ deviation away from the SM. 
This result bolsters the anomalous measurement of 2021~\cite{Muong-2:2021ojo, Girotti:2021vyl}, $ \Delta a_{\mu} = a_{\mu}^{\rm exp} -a_{\mu}^{\rm SM} = (251 \pm 59) \times 10^{-11}$  with somewhat greater uncertainity than the most recent one. That is why the most recent more precise measurement brings this aspect to probe new physics, back to the limelight. 
This result is also a confirmation of another previous similar measurement for $(g-2)_{\mu}$ in Brookhaven experimental observation~\cite{Muong-2:2006rrc}. 
%
%{\color{blue} Recently $(g-2)_\mu$ measurement at the Fermilab experiment has reported a 5.1 $\sigma$ deviation from the SM prediction of $\Delta a_{\mu}$ as~\cite{}  $$\Delta a_\mu^{\rm (FNAL + BNL)_{2023}} = 249(48) \times 10^{-11}$$
%}
 %More recently in August 2023, the FNAL came up with an even more precise result, with the uncertainty reduced by almost half, quoting a combined $\Delta a_\mu^{\rm (FNAL + BNL)_{2023}}$ = $249(48) \times 10^{-11}$, which is 5.1$\sigma$ away from the SM value considered here[will put ref]
The discrepancy between the SM and the measured value of  $(g-2)_\mu$ can be attributed to some new physics.

The addition of $U(1)_X$ gauge symmetry to the SM gauge group is one the simplest extensions of SM. 
The leptophilic models $U(1)_{L_\mu-L_\tau}$ and $U(1)_{L_e -L_{\mu}}$ are very interesting in this regard as they are the minimal anomaly-free U(1) models which can contribute to $(g-2)_\mu$. 
Among these two, only $U(1)_{L_\mu-L_\tau}$ model has parameter space allowed by recent experiments which can explain $(g-2)_\mu$~\cite{Bauer:2018onh}. 
This motivates us to consider the $U(1)_{L_\mu-L_\tau}$ model for our analysis. 
The additional gauge boson $Z'$ can contribute to  $(g-2)_\mu$ through its interaction with leptons.  Further, we can incorporate a vectorlike Dirac fermion DM candidate ($\chi$) in the model without any additional contribution to the gauge anomaly. 
The $Z^{\prime}$ can act as a possible mediator in DM annihilation. 
Different dark matter phenomenological implications of the $(g-2)_{\mu}$ observation in this setup were discussed already in the literature in Refs.~\cite{Qi:2021rhh, Hapitas:2021ilr, Gninenko:2018tlp, Borah:2021jzu, Borah:2021khc, Ko:2021lpx, Holst:2021lzm, Baek:2015fea, Patra:2016shz,Foldenauer:2018zrz,Costa:2022oaa}. 
But those studies were based on the previous $(g-2)_\mu$ results. In this article, we study the DM phenomenology for a GeV scale DM candidate in the light of the most recent $(g-2)_\mu$ \cite{Muong-2:2023cdq} measurement.

 Another interesting aspect of $U(1)_{L_\mu-L_\tau}$ model is that the neutrino floor, the irreducible background arising from the neutrino-nucleus interaction,  gets significantly enhanced in certain regions of parameter space \cite{Sadhukhan:2020etu}. Though the enhancement of the neutrino floor doesn't serve as an additional constraint in terms of ruling out the parameter regions, it can surely impact the detection prospects of the DM as it becomes impossible to distinguish between DM and the neutrinos in the current experimental facilities if the DM direct-detection cross-section resides within the neutrino-floor. Thus with the current set of direct-detection experiments, it is desirable to obtain parameter regions outside the neutrino floor while looking for prospects of the DM candidate, thereby constraining the $U(1)_{L_\mu-L_\tau}$. 

Keeping these aspects in mind, we set out to pursue a modest goal where we find out the prospects of a GeV scale DM in $U(1)_{L_\mu-L_\tau}$ by including the constraints coming from $(g-2)_\mu$ and the enhancement of neutrino floor. 
We also explicitly look at the importance of s and t channel annihilation channels in different regions of the parameter space and make a detailed analysis to show how the presence of $(g-2)_\mu$ constraint preferentially selects only a particular type of annihilation (t-channel ones) to satisfy the correct relic density. 
The conclusions regarding the direct detection prospect of the fermionic DM in this model gets the most significant modification in terms of the exact DM mass range that is allowed with or without the anomalous $(g-2)$ constraint. 
We also observe that combined constraints coming from DM direct detection and $(g-2)_\mu$ happen to be more constraining compared to the indirect detection constraints.

The rest of the article is written as follows:
In section~\ref{model} the $U(1)_{L_{\mu}-L_{\tau}}$ model is discussed by augmenting it with a vectorlike fermionic DM candidate. 
The existing constraints on the model are also listed and the parameter region satisfying the $(g-2)_{\mu}$ is also exhibited. 
The next section~\ref{nufloor} is dedicated to the neutrino floor computation, where the modification of the neutrino floor is pointed out in the parameter regions inspired by $(g-2)_{\mu}$. 
In section~\ref{relic-g2}, we discuss relic density constraints on the GeV scale DM. 
In section~\ref{dd-g2}, we impose further constraints coming from the DM direct detection experiments and the neutrino floor to study the effects of $(g-2)_{\mu}$ constraints on the existing parameter space. 
Then we briefly discuss the viability of indirect detection constraints in section~\ref{indirect}. 
Finally, in section~\ref{summary}, we summarize our results and chart out the possible road ahead.

%%%%%%%%%%%%%%%%%%%%%%%%%%%%%%%%%%%%%%%%%%%%%%
\section{Muon $g-2$ in $U(1)_{L_{\mu}- L_{\tau}}$}
\label{model}
We first discuss the $U(1)_{L_{\mu}- L_{\tau}}$ model briefly outlining the constraints on this prior to the muon $(g-2)$ anomalous measurement. Then it is explored how the inclusion of this new constraint can restrict the model parameter space. 
\vspace{-0.7cm}
\subsection*{$U(1)_{L_{\mu}- L_{\tau}}$ Model: Constraints and DM}
 In this article we consider the $U(1)_{L_{\mu}- L_{\tau}}$ model to study the feasibility of simultaneously explaining (g-2)$_\mu$, dark matter relic density and modification in neutrino floor. The $U(1)_{L_{\mu}- L_{\tau}}$ is one of the three simplest U(1) extension in SM that does not induce any extra gauge anomalies~\citep{Foot:1990mn,He:1990pn,He:1991qd,Choudhury:2020cpm}. The other two models being $U(1)_{L_{e}- L_{\mu}}$ and $U(1)_{L_{e}- L_{\tau}}$.
 
 In $U(1)_{L_{\mu}- L_{\tau}}$ model, the new U(1) symmetry can be broken spontaneously by a new complex scalar S. The symmetry breaking causes the $U(1)$ associated vector field $Z'$ to acquire mass through its interaction with S. The new Lagrangian terms are given by,
 \begin{eqnarray}
L_{new}=&& -\frac{1}{4} Z'^{\mu\nu}Z'_{\mu\nu} + \sum_l \bar{l}\gamma^\mu\left(-g_{\mu-\tau}\,Y'_l\, Z'_\mu\right)l \nonumber
\\
&& + \left(D_\mu S\right)^\dagger\left(D^\mu S\right)+\mu^2_S S^\dagger S + \lambda_{S}\left(S^\dagger S\right)^2\nonumber
\\
&&  + \lambda_{SH}\left(S^\dagger S\right)H^\dagger H
\end{eqnarray}
Here $\mu^2_S$ and $\lambda_S$ are bilinear and quartic self-interactions for S respectively. S couples with the SM Higgs $H$ via quartic coupling $\lambda_{SH}$. The $Z'$ boson interacts with fermions by total coupling $g_{\mu-\tau}Y'_l = g_{\mu-\tau}\left(L_\mu-L_\tau\right)$. 

\begin{figure}[h!]
\centering 
\begin{subfigure}{.4\textwidth}\centering
\includegraphics[scale=0.4]{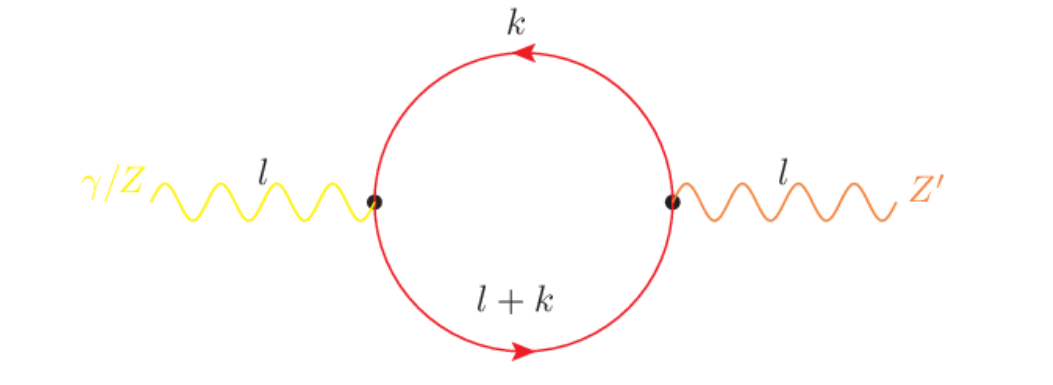}
%\caption{}
\end{subfigure}%
\caption{Lepton loop through which $Z^{\prime}-Z/\gamma$ mixing is induced in $U(1)_{L_\mu - L_\tau}$  model.} \label{feyn12}
\end{figure}

In this model, $Z'$ does not have tree-level interactions with quarks and electrons. The interactions are mediated by $Z/\gamma-Z^{\prime}$ loop given by the diagram in Fig.~\ref{feyn12}.

\begin{eqnarray}
\delta^{\mu\nu}_{ij}=&&\frac{1}{(2\pi^2)}\left[- l^\mu l^\nu + g^{\mu\nu} l^2\right]\nonumber
\\
&&\times\int^1_0 dx\,\left( \log\frac{x(x-1)l^2 + m^2_{l_i}}{x(x-1)l^2 + m^2_{l_j}}\right)x(1-x)
\end{eqnarray}
where, $l$ is momentum transfer, $m_{l_{i(j)}}$ is mass of i(j)$^{th}$ flavour lepton in  the loop. 

The model can incorporate a vector-like fermion (VLF), $\chi$, without introducing an extra gauge anomaly. The DM Lagrangian can take the form:
 \begin{eqnarray}
L_{DM}=&&  \bar{\chi}\left(i\gamma^\mu D_\mu -m\right)\chi . \nonumber
\end{eqnarray}
The $Z'$ also acquires interaction with the DM candidate $\chi$ through the covariant derivative $D_\mu\chi= \left(\partial_\mu -ig_\chi \right)\chi$ present in the DM kinetic term. The interactions between $Z'$ and fermions, dark or otherwise are highlighted by the interaction terms. 
\begin{eqnarray}
\label{intLij}
L_{i - f} &=& - g_{\mu-\tau}( \bar l_i \gamma^\mu l_i - \bar l_j \gamma^\mu l_j
+ \bar \nu_i \gamma^\mu L \nu_i - \bar \nu_j \gamma^\mu L \nu_j) Z^\prime_\mu\nonumber
\\
&&- g_\chi\bar\chi\gamma^\mu\chi\nonumber.
\end{eqnarray}
Here $ g_\chi$ is the total gauge charge of VLF under U(1)$_{L_\mu -L_\tau}$.
As $Z'$ interacts with both DM candidate $\chi$ and leptons: $\mu$ and $\tau$, it effectively acts as a portal between dark matter and SM particles. 
Therefore DM candidate, $\chi$ contributes to dark matter relic density through s-channel annihilation process $\bar{\chi} \,\chi \rightarrow Z'\rightarrow \bar{l}\,\,l\,(\bar{\nu}_l\,\, \nu) $ and t-channel process  $\bar{\chi} \,\chi \rightarrow Z'\,\,Z' $. 
For $m_\chi \leq M_{z'}$ s-channel is the dominant contribution to DM relic density whereas for $m_\chi \geq M_{z'}$ t-channel remains the dominant contribution. 
The annihilation can be detected through final state radiation (FSR) in the form of photon flux in Fermi-LAT experiment~\cite{Albert_2017}. 
Also, $\mu$ and $\tau$ leptons along with their anti-particles in final states can further decay to electron and positron before reaching Earth. 
Electron and positron excess have been measured at DAMPE~\cite{TheDAMPE:2017dtc} and AMS02~\cite{PhysRevLett.110.141102} experiments respectively. %We shall refrain from discussing the indirect detection of dark matter in detail. We shall discuss the direct detection of dark matter through a recoil of the target nucleus in detail in the following sections.

We refer to Ref.~\citep{Bauer:2018onh,Chun:2018ibr} for detailed understanding of $U(1)_{L_{\mu}- L_{\tau}}$ model constraints. 
In the electron beam dump experiments~\citep{Riordan:1987aw, Bjorken:1988as, Bross:1989mp} and the proton beam dump experiments~\citep{DORENBOSCH1986473, LSND:1997vqj, Blumlein:2011mv}, an accelerated beam of electrons and protons respectively falls on a target material. 
Secluded photons can be produced with electrons and protons as final state radiations and through the decay of mesons which are produced when energetic protons hit the target material. 
As the $Z'$ interactions with electrons and quarks are loop suppressed in this model, electron and proton beam dump constraints are weaker. 
TEXONO~\citep{Deniz:2009mu} experiment measures the elastic scattering of $\bar{\nu}_e$ produced at Kuo-sheng Nuclear power reactor with CsI(Tl) crystal array target. 
As the incident neutrino is $\bar{\nu}_e$, TEXONO experimental constraints are diminished for the model. 
Experiments like CCFR~\citep{CCFR:1991lpl}, Charm-II \citep{Vilain:1994qy} provide constraints on the model derived from the neutrino trident process $\nu_{\mu} Z \to \nu_{\mu} \mu^{+} \mu^{-} $. 
Borexino~\citep{Bellini:2011rx} measures the scattering of solar neutrinos with a liquid scintillator. Borexino and  Charm-II provide us the most stringent constraints in the region of interest, i.e. for $M_{z'}$ range 1-200 MeV. 
Currently, COHERENT experiment CE$\nu$NS measurements are preliminary and constraints remain comparatively weaker. 
In the future, with increased exposure COHERENT experiment will be able to probe the currently available parameter space for $(g-2)_{\mu}$. 
In addition, constraints derived from astrophysical observations and meson decays for an ultralight $Z^{\prime}$ ($M_{z^{\prime}} \leq 1 $\,eV),  have been studied in Ref.~\cite{Dror:2020fbh}.

\subsection*{Muon $g-2$}
\begin{center}
\begin{figure}[h!]
\includegraphics[scale=0.5]{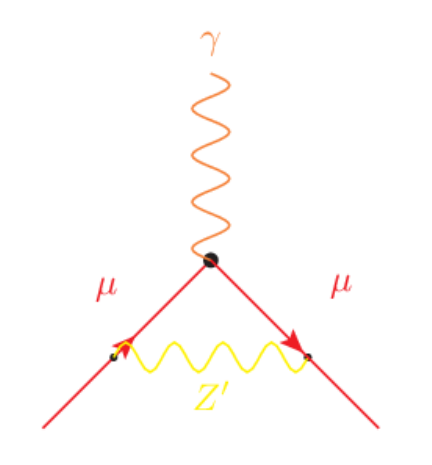}
\caption{\textit{BSM contribution to the muon (g-2) through a $Z^{\prime}$ mediated loop diagram.}}
\label{muon-g2}
\vspace{-0.3cm}
\end{figure}
\end{center}
\vspace{-0.3cm}
In $U(1)_{L_{\mu}- L_{\tau}}$ model, we get a contribution to muon-($g-2$) at one loop level due to the additional $Z'$ boson mediated diagram, which is given by:
\begin{equation}
\Delta a_\mu = \frac{g_{\mu-\tau}^2}{8 \pi^2} \int_{0}^{1} dx \frac{2 m_\mu^2 x^2(1-x)}{x^2 m_\mu^2 + (1-x)M_{z'}^2}.
\end{equation} 
There is a bit of tension in the theoretical prediction of muon $g-2$ within the SM. 
The SM value obtained by a data-driven approach using dispersion relation techniques ~\cite{Aoyama:2020ynm} shows a much higher deviation from the measured value than when the SM value is obtained from lattice results~\cite{Borsanyi:2020mff}. 
%{\color{green}The lattice results, however, are very recent and require more literature to confirm it in order to consider it with confidence.} 
Though the lattice computations are somewhat new, more work in that direction is required to support those before they can be taken into serious consideration.
We, therefore, stick to the data-driven theoretical result (which utilizes perturbative QCD) to carry out our analysis. 
With this result, the Brookhaven National laboratory(BNL) E821 experiment had earlier shown a  $\Delta a_\mu^{\rm BNL}$ = $279(76) \times 10^{-11}$~\cite{Muong-2:2006rrc}, thus resulting in a $\sim 3.7 \,\sigma$ deviation of from the SM value. 
Fermilab National Laboratory(FNAL) in 2021 confirmed the BNL results by reporting a deviation of $\Delta a_\mu^{\rm FNAL}$ = $230(69) \times 10^{-11}$. 
These two results had thus put a combined discrepancy of 4.2$\,\sigma$ from the SM with $\Delta a_\mu^{\rm (FNAL + BNL)_{2021}}$ = $251(59) \times 10^{-11}$~\cite{Muong-2:2021ojo, Girotti:2021vyl}. More recently in August 2023, the FNAL came up with an even more precise result (the most precise one up to date), with the uncertainty being reduced by almost half, providing a combined 
$\Delta a_\mu^{\rm (FNAL + BNL)_{2023}} = 249(48) \times 10^{-11},$ 
which is 5.1$\sigma$ away from the SM value considered here~\cite{Muong-2:2023cdq}. 
This result has generated a renewed interest in the anomalous $\Delta a_\mu$ related phenomenology, for which $U(1)_{L_{\mu}- L_{\tau}}$ model is a particularly suitable one, with increasing prospects for a light $Z'$ boson in the mass range $10-100$~MeVs. 
Figure~\ref{muon-g2} shows the allowed $\Delta a_\mu$ band at $2~\sigma$ confidence level as a function of the $M_{z'}$ mass and the coupling $g_{\mu-\tau}$ utilizing this combined BNL-FNAL result. 
\begin{center}
\begin{figure}[htb!]
\includegraphics[scale=0.4]{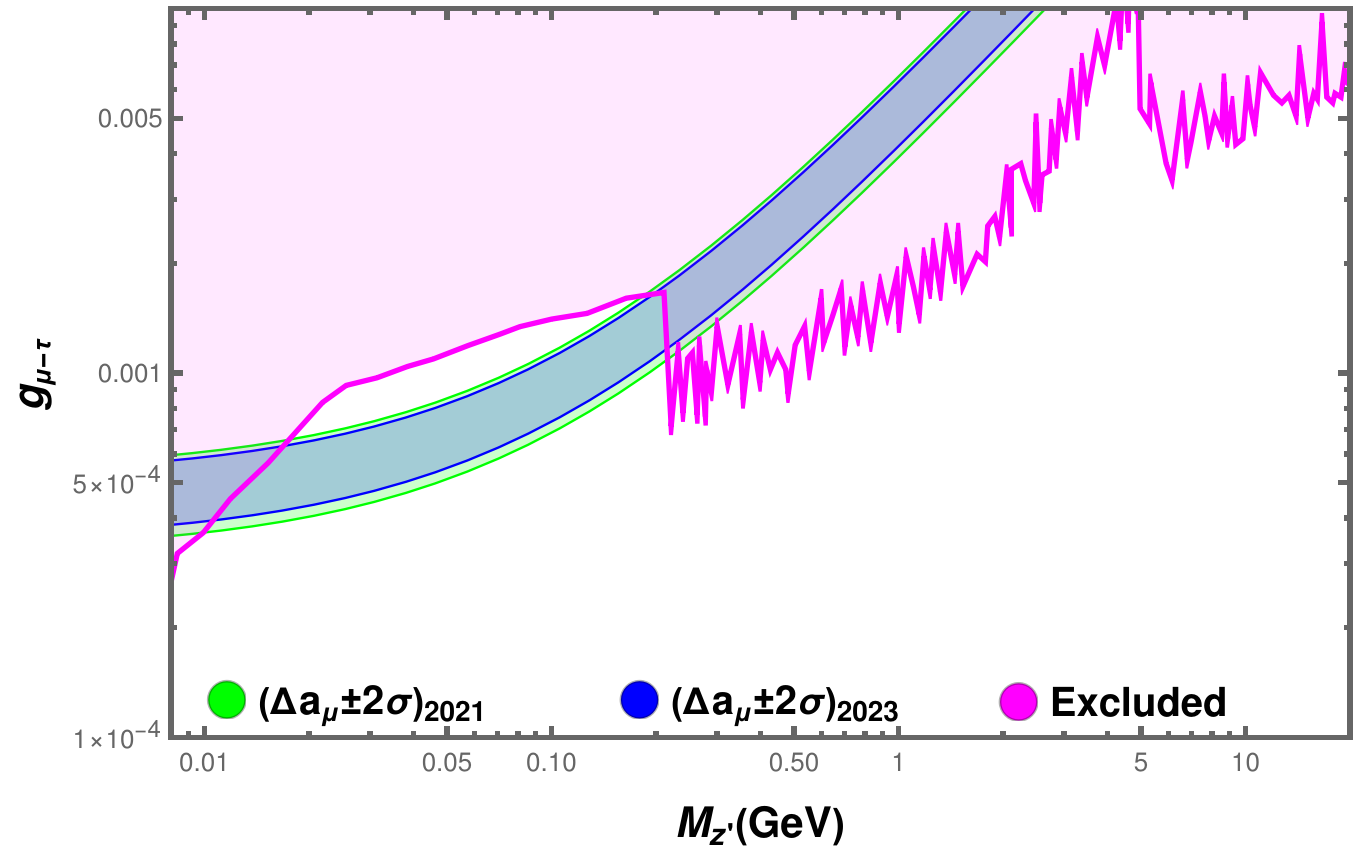}
\caption{\textit{Allowed parameter space for muon $g-2$.}}
\label{muon-g2}
\end{figure}
\end{center}
\vspace{-0.5cm}
For completeness, if we consider the lattice results for the SM value, we get a  $\Delta a_\mu^{\rm LO-HVP}(\rm lattice)$ =  $105(59) \times 10^{-11}$. 
If this result is indeed true, the discrepancy of the combined BNL-FNAL result with the SM comes within $2\, \sigma$. 
This can potentially reduce the motivation to probe new physics models.

\section{Neutrino Floor Modification: $(g-2)_{\mu}$ Effects} 
\label{nufloor}
As we discussed previously, in $U(1)_{L_\mu - L_\tau}$ model, the additional $Z'$ boson interacts with $\nu_\mu$ and $\nu_\tau$. Therefore non-standard coherent elastic neutrino-nucleus scattering (CE$\nu$NS) can be induced in the model through $Z'-\gamma$ mixing. 
\subsection*{Differential scattering cross-section}
First, we look at the case where astrophysical neutrinos are hitting the DM search detector material causing a nuclear recoil. Total neutrino-nucleus differential scattering cross-section in  $U(1)_{L_\mu - L_\tau}$ is given by,
\begin{widetext}
 \begin{eqnarray}
 \label{differentialscattering}
\frac{d\sigma_{\mu-\tau}}{dE_r}=&&\frac{d\sigma_{SM}}{dE_r}
-\frac{{m_N}\,G_f\, Q_{\nu N{\mu-\tau}} Q_{\nu N} \left(1-\frac{{E_r} {m_N}}{2 {E_\nu}^2}\right)F^2(E_r)}{\sqrt{2} \pi  \left(2 {E_r} {m_N}+M_{z'}^2\right)}+\frac{{m_N}\, Q_{\nu N{\mu-\tau}}^2 \left(1-\frac{{E_r} {m_N}}{2 {E_\nu}^2}\right)F^2(E_r)}{2 \pi  \left(2 {E_r} {m_N}+M_{z'}^2\right)^2},
\end{eqnarray}
\end{widetext}
where $\frac{d\sigma_{SM}}{dE_r}$ is the contribution from SM, given as, 
\begin{eqnarray}
&&\frac{d\sigma_{SM}}{dE_r}=G_f^2 \frac{m_N}{4 \pi } Q^2_{\nu N} \left(1-\frac{E_r m_N}{2 {E_\nu}^2}\right)F^2(E_r).
\end{eqnarray} 
Here $G_f$ is the Fermi constant, $Q_{\nu N} = N - (1-4\sin^2\theta_w) Z$ is effective weak hyper-charge in the SM for the target nucleus with $N$ neutrons and $Z$ protons and $F(E_r)$ is the Helm form factor (Ref.~\cite{Lewin:1995rx}). The effective coupling between neutrino and nucleus for  $U(1)_{L_\mu - L_\tau}$ model can be written as,
\begin{eqnarray}
 Q_{\nu N{\mu-\tau}} =  g_{\mu-\tau}^2\frac{2\, \alpha_{EM}}{\pi} \delta_{\mu\tau} Z 
\end{eqnarray}
where $ g_{\mu-\tau}$ is coupling given in Eq.~\ref{intLij},  $\alpha_{EM}$ is the fine structure constant and $\delta_{\mu\tau}$ is the scalar part of loop factor given in equation \ref{intLij}. 
\subsection*{Neutrino and Dark Matter Rate equation}
%\subsubsection*{neutrino rate equation}
%Using the differential cross-section of the neutrino scattering and augmenting that with the source and detector effects, the scattering rates can be obtained. 
CE$\nu$NS rate for a detector arising from astrophysical neutrino flux is given by,
\begin{eqnarray}
\label{ratenu}
\frac{d R_{\nu-N}}{d E_r} =&& \frac{\epsilon}{m_N} \int_{E^{min}_\nu} \mathcal{A}(E_r) \left.\frac{d \phi_\nu}{d E_{\nu}}\right|_{\nu_\alpha} P(\nu_\alpha \rightarrow\nu_\beta,E_\nu)\nonumber
\\
&&\times \frac{d \sigma (E_\nu,E_r,\nu_\beta)}{d E_r} d E_\nu
\end{eqnarray}
Here $\epsilon$ is the exposure of the experiment measured in units of mass $\times$ time, $\mathcal{A}(E_r)$ is the detector efficiency and is set to one in the following calculations. $E^{min}_\nu$ is the minimum incident neutrino energy required to produce a detectable recoil for a material nucleus of mass $m_N$ with energy $E_r$, which in the limit of $m_N >> E_\nu$ can be written as,
\begin{eqnarray}
E^{\rm min}_\nu = \sqrt{\frac{m_N E_r}{2}}.
\end{eqnarray}

Here, $\frac{d \sigma (E_\nu, E_r, \nu_\beta)}{d E_r}$ is $\beta$ flavor dependent neutrino-nucleus differential scattering cross-section and $\left.\frac{d \phi_\nu}{d E_{\nu}}\right|_{\nu_\alpha}$ is  the incoming  neutrino flux of flavor $\alpha$. The fluxes used in this analysis involve fluxes from solar, atmospheric, and diffuse supernova neutrinos, which can be found in  Refs.~\cite{Strigari:2009bq,Billard:2013qya}.

\subsubsection*{Dark matter direct detection rate equation}
For the DM, the spin-independent DM-nucleon scattering rate equation is given by,
\begin{equation}
\label{rateDM}
\frac{d R_{ DM-N}}{d E_r} = \epsilon\,\frac{\rho_{DM}\sigma^0_n A^2}{2m_{DM}\mu^2_{n}}F^2(E_r) \int_{v_{min}} \frac{f(v)}{v } d^3 v
\end{equation}
Here $\epsilon$ is the exposure of the detector given in units of MT (mass$\times$time), $m_{DM}$ is the DM mass $\mu_{n}$ is DM-nucleon reduced mass, $A$ is the mass number of target nuclei, $\sigma^0_n$ is the DM-nucleon scattering cross-section at zero momentum transfer. $F(E_r)$ is the Helmholtz form factor. $\rho_{DM}$ is the local density of the DM in the immediate vicinity of the Earth and f(v) the is DM velocity distribution in the Earth’s frame of
reference, which for the case of Maxwell distribution is given by,
\begin{eqnarray}
\int_{v_{min}} \frac{f(v)}{v}d^3 v =&& \frac{1}{2v_0 \eta_{E}}\left[erf(\eta_+)-erf(\eta_-)\right]\nonumber
\\
&&- \frac{1}{\pi v_0 \eta_E }\left(\eta_+-\eta_-\right)e^{\eta^2_{esc}}
\end{eqnarray}
Here $\eta_E$=$\frac{v_E}{v_0}$, $\eta_{esc}$=$\frac{v_{esc}}{v_0}$ and $\eta_\pm$ = $min\left(\frac{v_{min}}{v_0}\pm \eta_E,\frac{v_{esc}}{v_0}\right)$, where $v_0$ is local galactic rotational velocity, $v_{E}$ velocity of Earth with respect to galactic center, $v_{esc}$ escape velocity of DM from galaxy. We have used values  $v_0=220$km/s, $v_E=$232 km/s and $v_{esc}=$ 544 km/s  in above calculations. %\cite{Barger:2010gv,Chao:2019pyh}
\subsection*{Neutrino Floor}

The Neutrino floor is defined as the value of the DM-nucleon scattering cross-section at which the ratio of the number of scattering events generated by DM-nucleon scattering to that of events generated by neutrino background is 2.3 to 1 (at 90\% CL). This establishes a borderline below which the contribution of neutrino background increases and certainty that the observed events if any are from DM-nucleon scattering decreases.  
It is calculated by measuring the exposure required to observe a neutrino scattering event and then for the same exposure calculation the value of DM-nucleon scattering cross-section for which we get 2.3 scattering events (For more details see Ref.~\cite{Sadhukhan:2020etu} ).
\begin{eqnarray}
\label{Nflms}
\int^{E_{DM}^{max}}_{E_{th}}\frac{dR_{DM-N}}{dE_r}dE_r =   \frac{2.3}{1} \int^{E_r^{max}}_{E_{th}}\frac{dR}{dE_r}dE_r \nonumber ,
\end{eqnarray} 

 \begin{figure*}
\centering
\begin{subfigure}{.42\textwidth}\centering
  \includegraphics[width=\columnwidth]{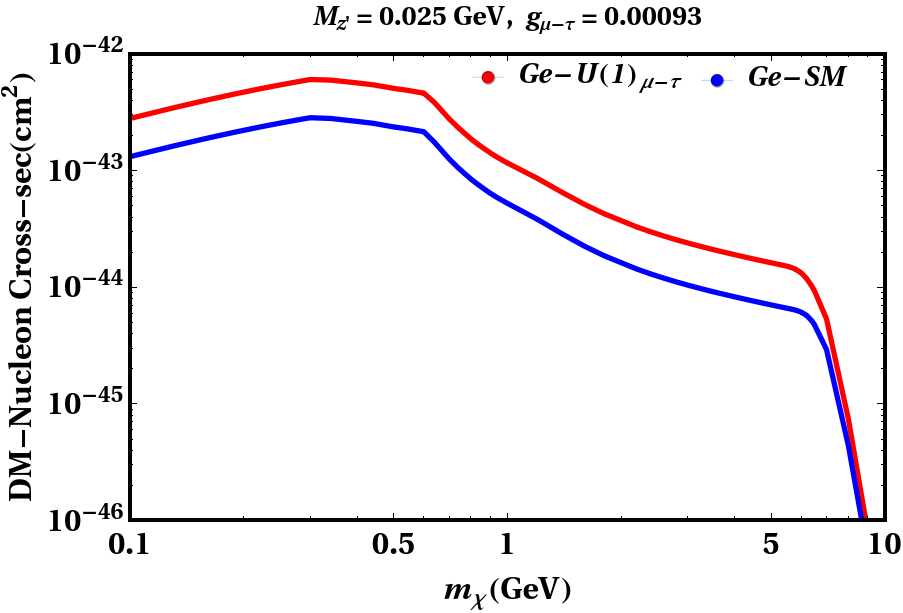}~~~~~~~~~~~
  \caption{$M_{z'}= 25$~MeV, $g_{\mu-\tau} = 9.3 \times 10^{-4}$. Maximum enhancement of the neutrino floor in the absence of $(g-2)_\mu$ constraint, for a Ge-based DM detector.}
  \label{NfloorGec}
\end{subfigure}%
\begin{subfigure}{.42\textwidth}\centering
  \includegraphics[width=\columnwidth]{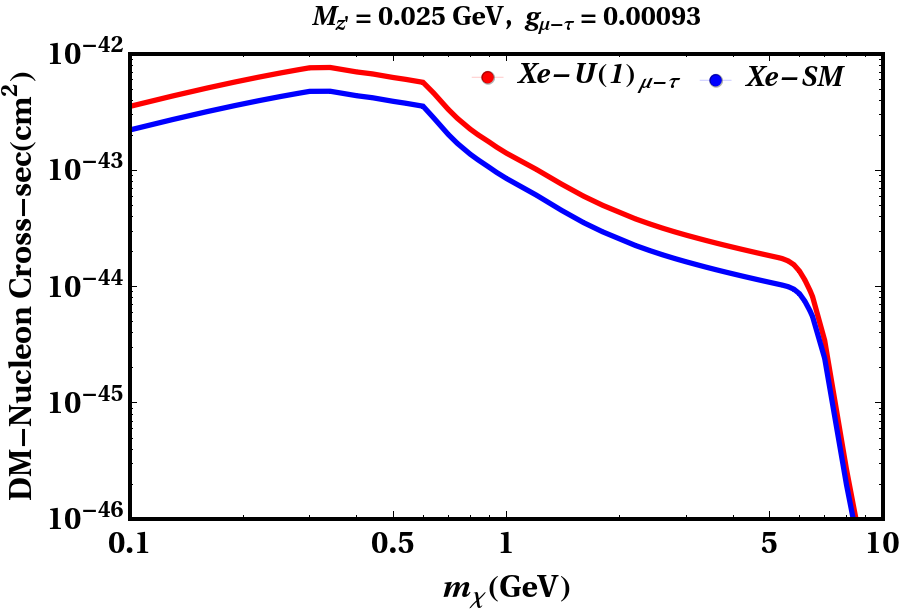}
 \caption{$M_{z'}= 25$~MeV, $g_{\mu-\tau} = 9.3 \times 10^{-4}$. Maximum enhancement of the neutrino floor in the absence of  $(g-2)_\mu$ constraint, for a Xe based DM detector.}
  \label{NfloorXec}
\end{subfigure}%
\\
\begin{subfigure}{.42\textwidth}\centering
  \includegraphics[width=\columnwidth]{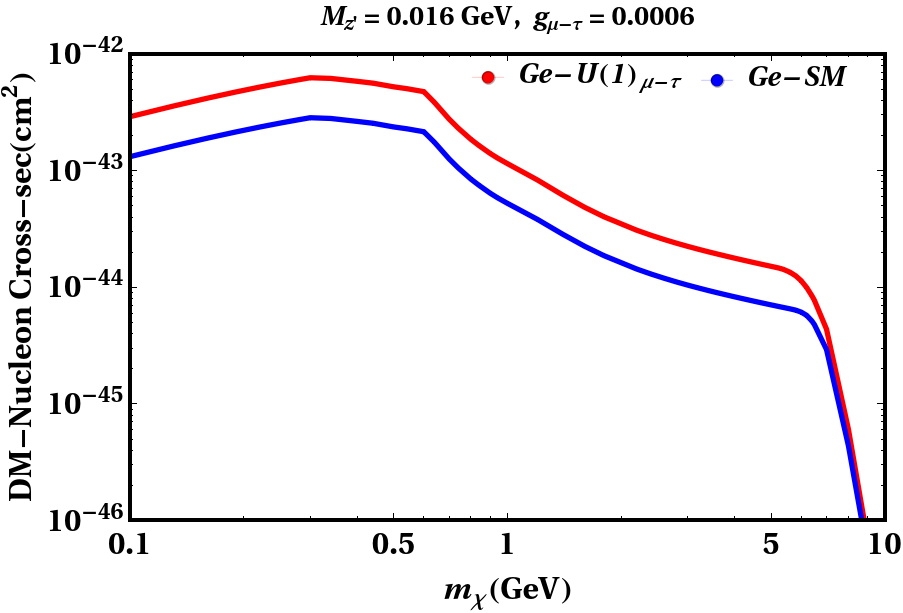}~~~~~~~~~~
  \caption{$M_{z'}= 16$~MeV, $g_{\mu-\tau} = 6 \times 10^{-4}$, when $(g-2)_\mu$ modification is taken into account. }
  \label{NfloorGea}
\end{subfigure}%
\begin{subfigure}{.42\textwidth}\centering
  \includegraphics[width=\columnwidth]{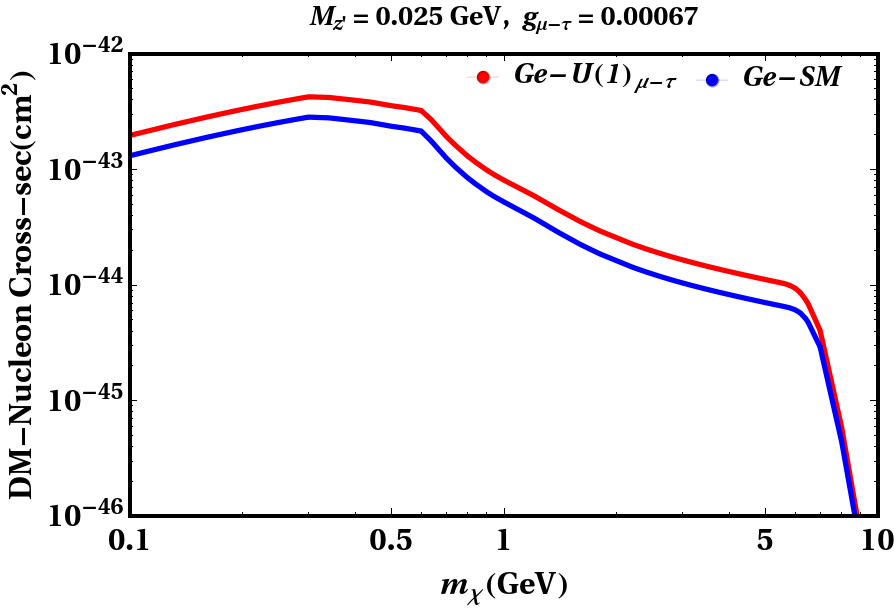}
 \caption{$M_{z'}= 25$~MeV, $g_{\mu-\tau} = 6.7 \times 10^{-4}$. Enhancement is shown after the inclusion of the $(g-2)_\mu$ constraint. }
  \label{NfloorXea}
\end{subfigure}%
\\
\caption{\small \em{Neutrino floor projected in the $\sigma^0_n$ vs $m_{DM}$ plane. Comparison of the neutrino floor for the SM (presented by the blue line) and that for  $U(1)_{L_\mu - L_\tau}$ (presented by the red line and labelled as  $U(1)_{\mu - \tau}$). Figures show modifications for different detector materials, Ge$^{68}$  and Xe$^{131}$ .}}
\label{Nflr}
\end{figure*}
This can be reiterated in the form of the master equation,
\begin{widetext}
\begin{eqnarray}
\label{Nflcalc}
\sigma^0_n =&&  \frac{2.3}{1}\left(\int^{E_{max}}_{E_{th}}\frac{1}{m_N}\int_{E^{min}_\nu} \left.\frac{d \phi_\nu}{d E_{\nu}}\right|_{\nu_\alpha} P(\nu_\alpha \nu_\beta,E_\nu) \frac{d \sigma (E_\nu,E_r,\nu_\beta)}{d E_r} d E_\nu\right)\left(\frac{\rho_{DM} A^2}{2m_{DM}\mu^2_{n}}\int^{E^{max}_{DM}}_{E_{th}}F^2(E_r) \int_{v_{min}} \frac{f(v)}{v } d^3 v\right)^{-1}\nonumber
\\
\end{eqnarray}
\end{widetext}
Here $E^{max}_{DM}$ is the maximum recoil energy of DM with mass $m_{DM}$ can produce in a given nucleus. It is written as, 

$ 2\, m_{DM} \left(\frac{m_N m_{DM} } {(m_N + m_{DM})^2}\right) v_{esc}^2$.

Using the master equation~\ref{Nflcalc}, we evaluate the modified neutrino floor for the allowed region in Fig.~\ref{muon-g2}. 
It is noticed that enhancement in neutrino floor increases with increasing value of $M_{z'}$ till 25 MeV for the corresponding maximum value of allowed $g_{\mu-\tau}$ and starts to decrease with larger masses and becomes negligible around and after $M_{z'}$= 200 MeV. 
This can be understood from the profile of the excluded region seen in Fig.~\ref{muon-g2} which shows a sharp kink around $M_{z'}$ = 25 MeV. 
Also the beyond standard model contribution to CE$\nu$NS is proportional to $\frac{g^2_{\mu-\tau}}{2 m_N E_r + M_{z'}^2}$, therefore neutrino floor enhancement decreases with larger $M_{z'}$ values and lower $g_{\mu-\tau}$. 
We choose the benchmark points $M_{z'}=\{16, 25, 50, 200\}$ MeV to discuss the enhancement of the neutrino floor at different DM masses in tabulated form in the Appendix~\ref{tables}. 
As the chosen region is also important from the recent $(g-2)_\mu$ result perspective, we also tabulate neutrino floor enhancement for points with anomalous $(g-2)_\mu$ allowed couplings for the same benchmarks.

In Fig.~\ref{Nflr} we show graphs showing a modification of the theoretically estimated neutrino floor on the $\sigma^0_n-m_{DM}$ plane. 
Solid lines exhibit the boundary of the DM-nucleon scattering cross-section above which we can be certain (90\% C.L.) that the ratio of DM-scattering events is more than 2.3 times of the neutrino-scattering events. 
Blue contours signify neutrino floor estimated from Standard Model interactions whereas red contours show neutrino floor for $U(1)_{L_\mu - L_\tau}$ model. 
It can be noticed that modification in neutrino floor is most significant for DM mass below 7 GeV and diminishes to a barely noticeable change for higher masses. 
This can be understood as below 7 GeV DM mass $\sigma_n^0$ is sensitive to threshold recoil energies of 1 keV and less. 
Below 1 keV threshold energy CE$\nu$NS rate for the $U(1)_{L_\mu - L_\tau}$ model show enhancement of the same order as seen in neutrino floor below 7 GeV DM mass, and for threshold energy greater than 1 keV CE$\nu$NS rate modification in  $U(1)_{L_\mu - L_\tau}$ rapidly diminishes (For details see Ref.~\cite{Sadhukhan:2020etu}). 
The top panels Fig.~\ref{NfloorGec} (left) and Fig.~\ref{NfloorXec} (right) show neutrino floor contours for the benchmark point $M_{z'}= 25$ MeV and $g_{\mu-\tau} = 9.3 \times 10^{-4}$ for Germanium and Xenon based experiments respectively. 
This point corresponds to the maximum enhancement seen in the allowed parameter space. 
Modification by a factor of around 2.2 and 1.6 times are observed respectively in the neutrino floor below the DM mass around 7 GeV. 
Bottom left panel~\ref{NfloorGea} shows modification in neutrino floor for the benchmark point $M_{z'}= 16$ MeV, $g_{\mu-\tau}=6 \times 10^{-4}$ and bottom right panel~\ref{NfloorXea} shows modification in neutrino floor for the benchmark point $M_{z'}= 25$ MeV, $g_{\mu-\tau}=6.7 \times 10^{-4}$. 
These benchmark points are also within viable parameter space to explain the BNL-FNAL $(g-2)_{\mu}$ anomaly for  $U(1)_{L_\mu - L_\tau}$ model. 
Modification of around 2 and 1.6 times is seen below DM mass 7 GeV for Germanium and Xenon-based detectors respectively.
 %\newpage 
 %%%%%%%%%%%%%%%%
  %%%%%%%%%%%%%%%%
\section{GeV scale Dark Matter: Role of  $(g-2)_{\mu}$ Anomaly}
\label{relic-g2}
\begin{center}
\begin{figure}[h!]
\includegraphics[scale=0.55]{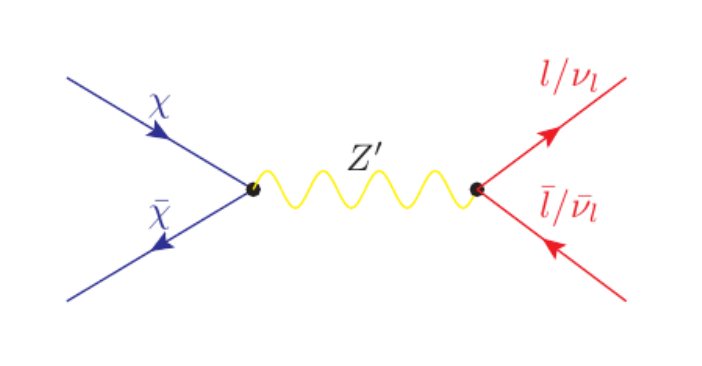}\hspace{0.8cm}
\includegraphics[scale=0.55]{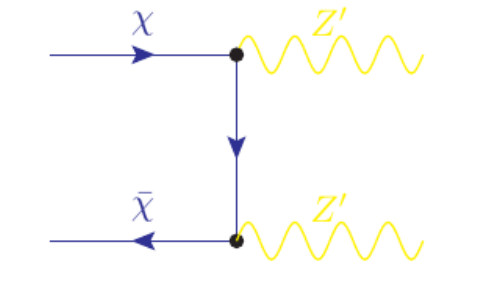}
\caption{\textit{The annihilation diagrams of DM candidates: The $Z^{\prime}$ mediated s-channel annihilation $\chi \chi \to \rm SM \,SM $ (top) and $Z^{\prime}$ mediated t-channel annihilation $\chi \chi \to Z^{\prime} Z^{\prime} $ (bottom) }}
\label{anni}
\end{figure}
\end{center}
\vspace{-0.4cm}
We have discussed before that for the DM, 100 MeV to 10 GeV mass scale is an interesting parameter space that is relatively less constrained from the experiments. 
Therefore, we can explore the possibility of a light DM candidate, mainly focusing on its direct detection. 
Here we work with a light vectorlike fermion as a DM candidate. 
At mass scales around 100 MeV, DM is not sufficiently heavy to contribute anything to cosmological observables as it cannot remain in thermal equilibrium at the energy scale of Big Bang Nucleosynthesis (BBN), thus contributing minimally to effective degrees of freedom~\cite{Drees:2021rsg}. 
On the other end of the spectrum, a 10 GeV dark matter is light enough to not get significantly constrained from the ongoing LHC searches, mainly through the jet/lepton plus missing energy (j/l + MET) channels. 
Our analysis can be extended to DM masses of around 10 MeV, but the detection mechanism is still not robust enough to have a conclusive opinion.
This motivates us to carefully probe the GeV scale DM territory which simultaneously satisfies the anomalous $(g-2)_{\mu}$, which has again got a boost in a recent result.

\begin{figure*}
\begin{center}
\includegraphics[scale=0.2]{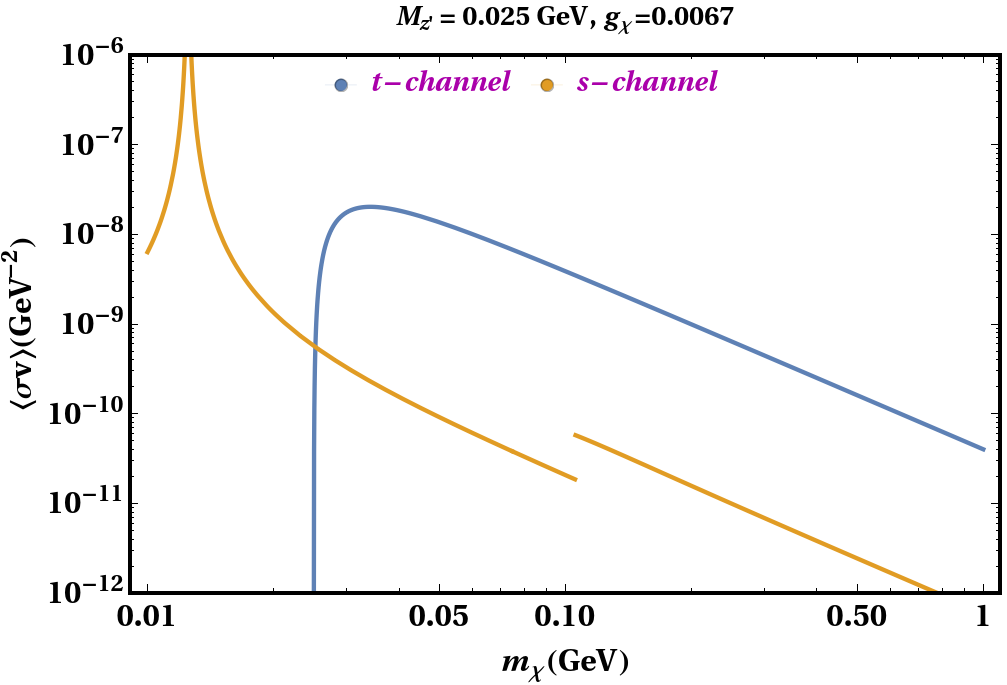}~~~~~
\includegraphics[scale=0.2]{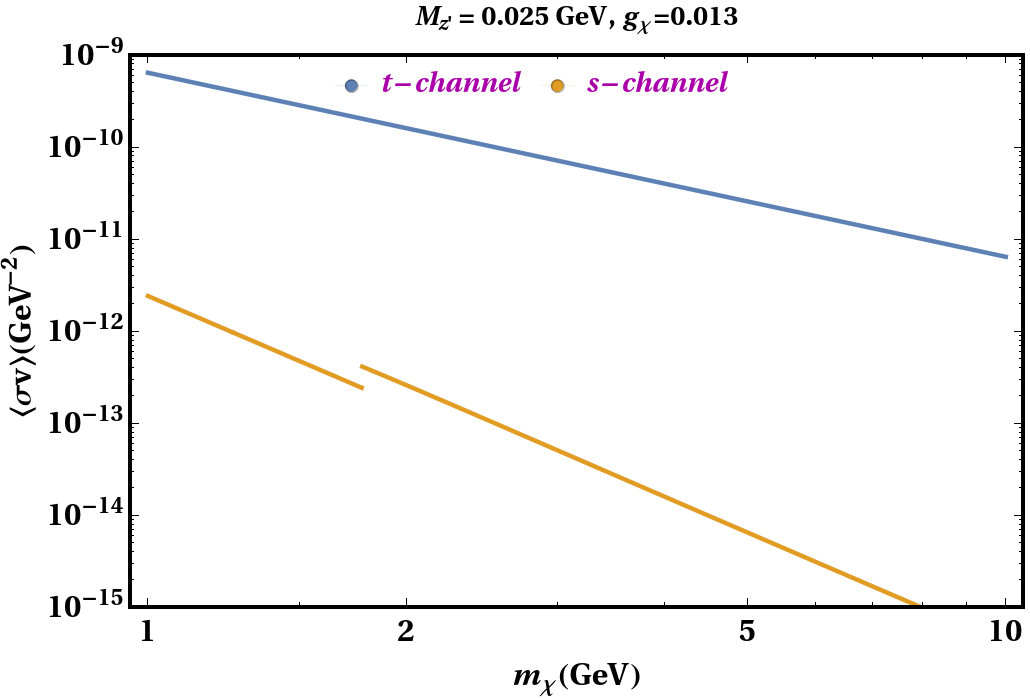}
\includegraphics[scale=0.2]{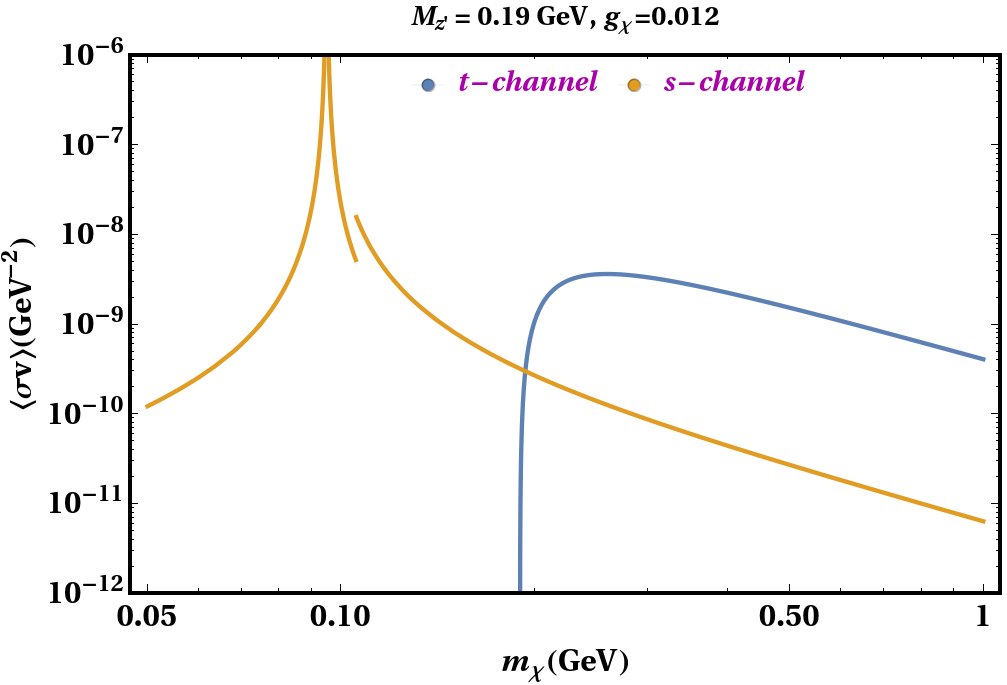}~~~~~
\includegraphics[scale=0.2]{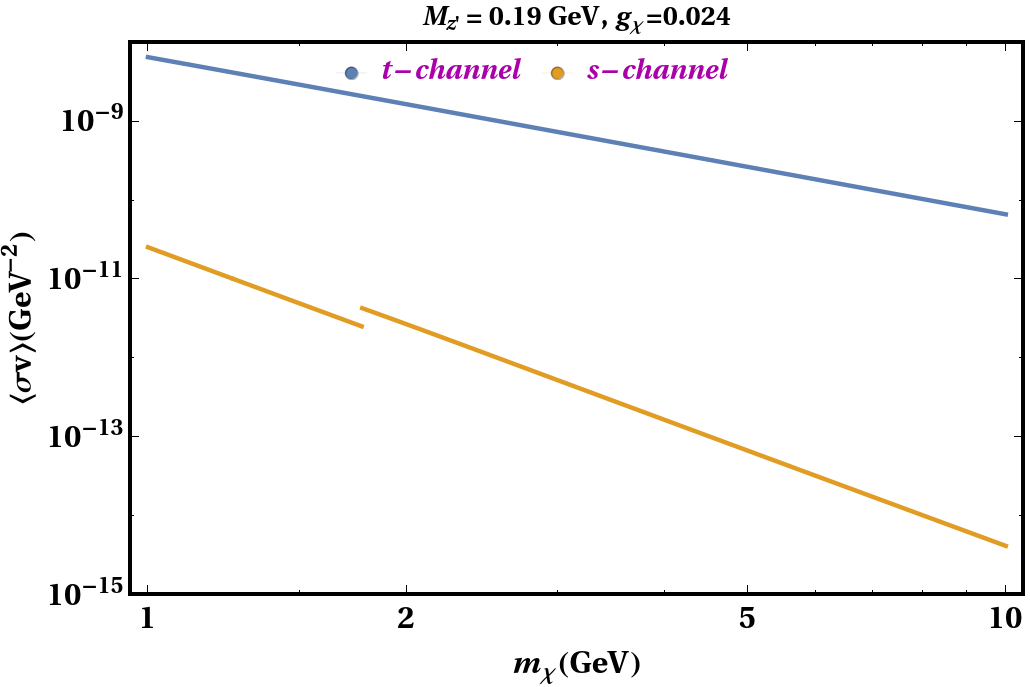}
\caption{\textit{Cross-section ($GeV^{-2}$) for the t-channel and s-channel annihilation diagrams}}
\label{annihilation-cx}
\end{center}
\end{figure*} 
\vspace{-0.4cm}
\subsection*{Relic Density Constraints: Role of $(g-2)_{\mu}$ Anomaly}
The Dirac fermion DM ($\chi$) introduced here has coupling with  $U(1)_{L_\mu - L_\tau}$ gauge fermion through an interaction term in the Lagrangian like 
$$ g_\chi \bar{\chi} \gamma^{\mu} Z^{\prime}_{\mu} \chi$$, 
where we take the effective $Z^{\prime}-\chi$ coupling as $ g_{\chi}$. 
The DM annihilation in this scenario happens through three major channels, namely $\chi \chi \to \ell \ell$ and $\chi \chi \to \nu \nu$, both through s-channel $Z^{\prime}$ mediated annihilation and a t-channel annihilation process $\chi \chi \to Z^{\prime} Z^{\prime}$, where the $Z^{\prime}$ can further decay to the SM muons and tau leptons. 
The annihilation processes are depicted in Fig.~\ref{anni}. 
The s-channel and t-channel annihilation cross sections are computed following the Re.~\cite{Altmannshofer:2016jzy}.
The s-channel diagram contributions to the DM annihilation:
\begin{align}
\langle \sigma v  \rangle (\chi \bar{\chi} \to \ell \bar{\ell}) \approx \frac{  g_{\chi}^2 g_{\mu - \tau}^2}{2 \pi} \sqrt{1- \frac{m_{\ell}^2}{m_{\chi}^2}}
 \frac{2 m_{\chi}^2 + m_{\ell}^2}{(4 m_{\chi}^2 - M_{z'}^2)^2},   \nonumber \\  
\langle \sigma v  \rangle (\chi \bar{\chi} \to \nu \bar{\nu}) \approx \frac{  g_{\chi}^2 g_{\mu - \tau}^2}{2 \pi } 
 \frac{ m_{\chi}^2 }{(4 m_{\chi}^2 - M_{z'}^2)^2},
 \end{align}
 The t-channel contribution is computed as 
 \begin{align} 
 \langle \sigma v  \rangle (\chi \bar{\chi} \to Z^{\prime} Z^{\prime}) \approx \frac{ g_{\chi}^4 }{16 \pi m_{\chi^2}} \left(1-\frac{M_{z'}^2}{m_{\chi}^2}\right)^{3/2} \left(1-\frac{M_{z'}^2}{2 m_{\chi}^2}\right)^{-2},
 \end{align}
 providing total thermally averaged annihilation cross section as $ \langle \sigma v  \rangle = \langle \sigma v  \rangle (\chi \bar{\chi} \to \ell \bar{\ell}) + \langle \sigma v  \rangle (\chi \bar{\chi} \to \nu \bar{\nu}) + \langle \sigma v  \rangle (\chi \bar{\chi} \to Z^{\prime} Z^{\prime})$. 
 
Initially, when the Universe was hot, DM particles were in thermal equilibrium with other SM particles inside the thermal plasma through $2 \to 2$ annihilation processes. 
As the Universe cools down as it expands, the DM annihilation does not remain efficient enough to keep them in thermal equilibrium as two DM particles cannot come together for annihilation. 
This is called the thermal freeze-out. 
After the freeze-out of the dark matter out of the thermal equilibrium, the remnant density of the dark matter that remains till today, the DM relic density is given as, 
\begin{align} 
\Omega h^2 \approx \frac{1.04 \times 10^9 x_F \rm GeV^{-1}}{\sqrt{g} M_{pl} \langle \sigma v \rangle},
\end{align}
where $x_F$ dictates the time or temperature of the thermal freeze-out given as $x_F = \frac{m_{\chi}}{T_F}$, with $T_F$ as the freeze-out temperature. 
For the freeze-out point fixation and DM relic density computation, we closely follow the approximate method we had outlined in Ref.~\cite{Borah:2017dqx}. 
For the GeV scale, Dirac fermionic DM, the freeze-out temperature is fixed at $x_F \sim (17-20)$, and this is obtained by solving the Boltzmann equation governing the thermal evolution of the DM.

  A DM in the mass range of $0.1-10$ GeV can annihilate resonantly if there is lighter $Z^{\prime}$ to assist through the s-channel process. 
However, if we impose the muon $(g-2)$ constraint, we are forced to stay in the $Z'$ mass range of 10-200 MeV. 
This implies that we are almost always away from the resonance peak for s-channel annihilation to be effective. 
Hence to simultaneously satisfy $(g-2)$ of the muon and the DM relic density, the t-channel plays the major role in DM annihilation. 
A comparison of s and t channel contributions is shown in Fig.~\ref{annihilation-cx} for two regions of our interest: 
I. sub-GeV dark matter ($m_{\chi}= 0.1-1 ~$GeV) 
II. super-GeV dark matter ($m_{\chi}= 1-10 ~$GeV). 
In the top two plots of Fig.~\ref{annihilation-cx}, we plot the t and s channel contribution to the annihilation cross section as a function of DM mass by fixing $M_{z'}$ at 25 MeV. 
We fix $g_{\chi}$ at $6.7 \times 10^{-4}$ for $m_\chi$ in 0.1-1 GeV and $g_\chi$ at $1.3 \times 10^{-2}$ for $m_\chi$ in 1-10 GeV. 
The choices of $g_\chi$ are dictated by the fact that for a thermal relic, we require a $\langle \sigma v \rangle$ around the range of $10^{-10}-10^{-9} ~ GeV^{-2}$. 
With the t-channel contribution being more important after the imposition of $(g-2)_{\mu}$ constraint, the annihilation cross-section decreases with increasing DM mass, thereby pushing the couplings to larger values to satisfy the correct relic density. 
As shown in the previous section, this is also incidentally the region where the enhancement of the neutrino floor is significant. 
Away from the resonance peak, the s-channel cross section is almost negligible compared to the t-channel one. 
The annihilation cross-sections increase slightly (both s and t) as we move towards a higher $Z'$ mass because the allowed $g_{\mu-\tau}$ from the muon $(g-2)$ band increases thereby resulting in higher cross-sections. 
This can be seen from the bottom two plots of Fig.~\ref{annihilation-cx} where the benchmark is chosen to be $M_{z'}$ of 190 MeV, with $g_\chi$ of $1.2 \times 10^{-2}$ for a lighter $\chi$ at masses in the range $0.1-1$~GeV while $g_\chi$ is fixed at $2.4 \times 10^{-2}$ at larger $m_\chi$ of $1-10$~GeV. 
Even though the neutrino-floor is not enhanced in this part of parameter space, this region has an impact on the direct-detection experiments, which we will see in the next section. 
One important point to note here is that earlier studies have only focussed on the resonant region where s-channel annihilation is dominant, but here we show that t-channel plays a very important role in sub-GeV DM annihilation being combined with the anomalous muon $(g-2)$. 
The price to pay is the slightly higher value of $g_\chi$ compared to $g_{\mu-\tau}$, but from a bottom-up phenomenological point of view, such values are still acceptable. 
This can also be somewhat redeemed by making the gauge charges of the muon and tau small fractional values, which will pull the allowed value of $g_{\mu-\tau}$ slightly higher, thereby reducing the hierarchy.
\vspace{-0.4cm}
 %%%%%%%%%%%%%%%%%%%%%%%%%%%%%%%%%%%%%
\section{DM direct detection vs Neutrino Floor: $(g-2)_{\mu}$ Effects}
\label{dd-g2}
\begin{center}
\begin{figure}[htb!]
\includegraphics[scale=0.3]{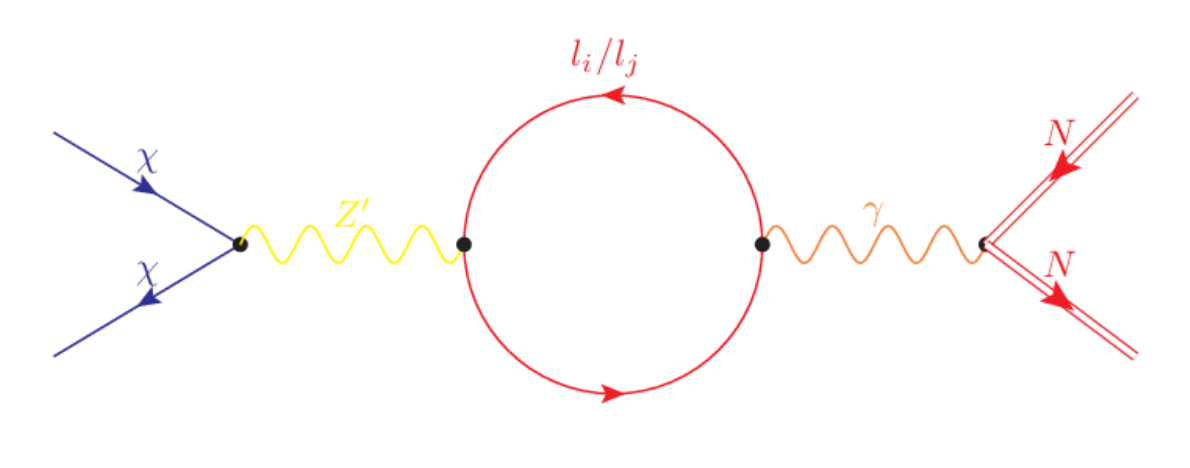}
\caption{\textit{DM direct detection process through DM-N scattering that can only happen in this theory through a loop diagram. }}
\label{DM_dd}
\end{figure}
\end{center}

The model in contention, being leptophilic, does not have any direct coupling between the DM particle and the quarks. 
Hence any scattering off the nuclei of the form $\chi N \rightarrow \chi N$ can only be rendered through a loop of charged leptons, where photons emitted by virtual leptons couple to the charge of the nucleons inside a nucleus as shown in Fig~\ref{DM_dd}.
 The expression for spin-independent cross section for the DM-Nucleon scattering can be written as:
\begin{equation}
\sigma_{SI}=\frac{\mu_N^2}{9 \pi A^2}\Big(\frac{\alpha_{EM} Z g_{\chi}g_{\mu-\tau} }{\pi M_{z'}^2 } \log \frac{m_\mu^2}{m_\tau^2}  \Big)^2
\end{equation}  
where $\mu_N$ is the reduced mass of the nucleon-DM system given by $m_N m_\chi/(m_N + m_\chi)$, with $m_N$ being the nuclear mass, $A$ and $Z$ being mass number and atomic number respectively. 
We stick to $Z=54$ and $m_N=129$~GeV consistent with Xenon (Xe) based experiments.

\begin{center}
\begin{figure}[!htb]
\includegraphics[scale=0.17]{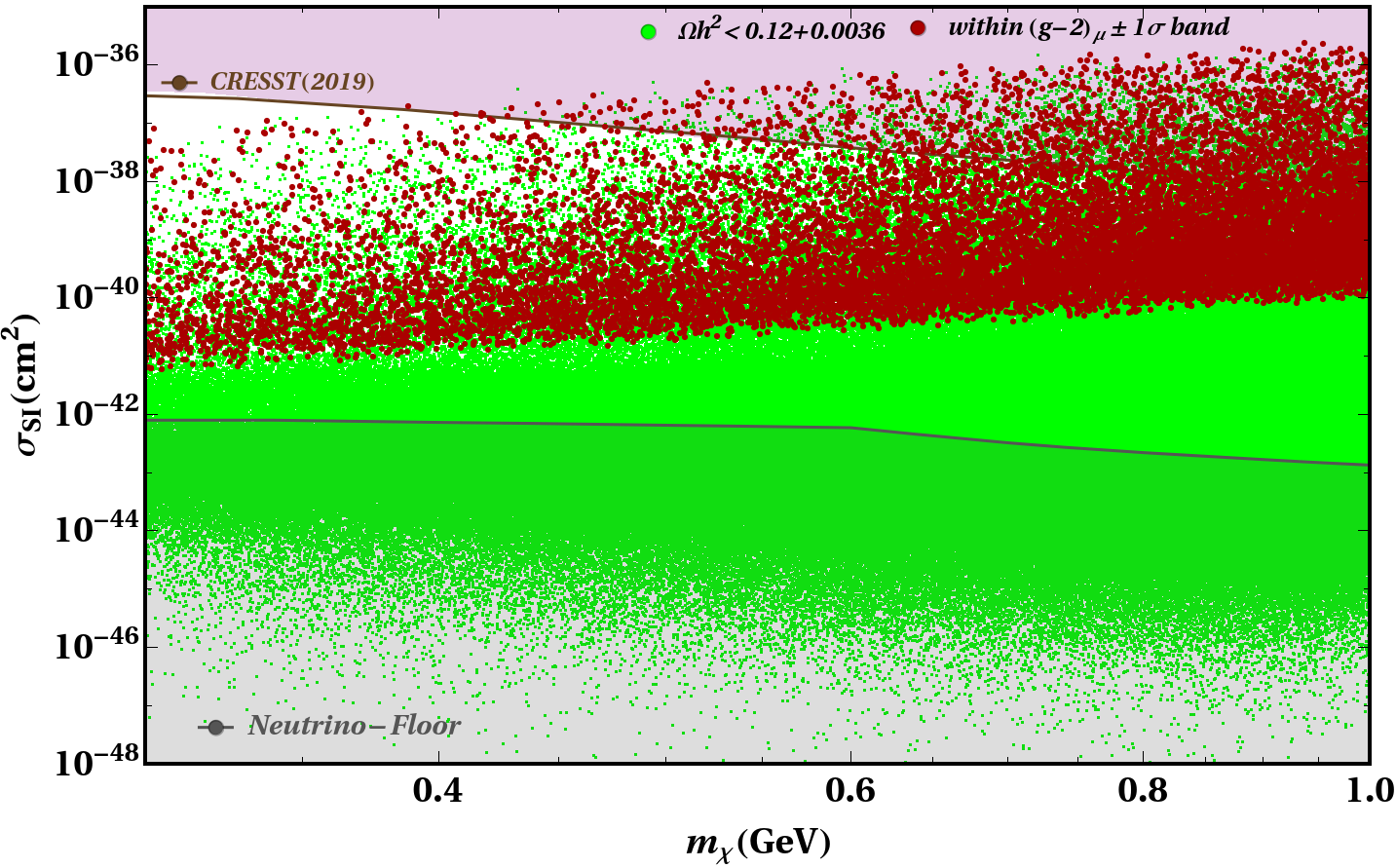}\vspace{0.2cm}
\includegraphics[scale=0.17]{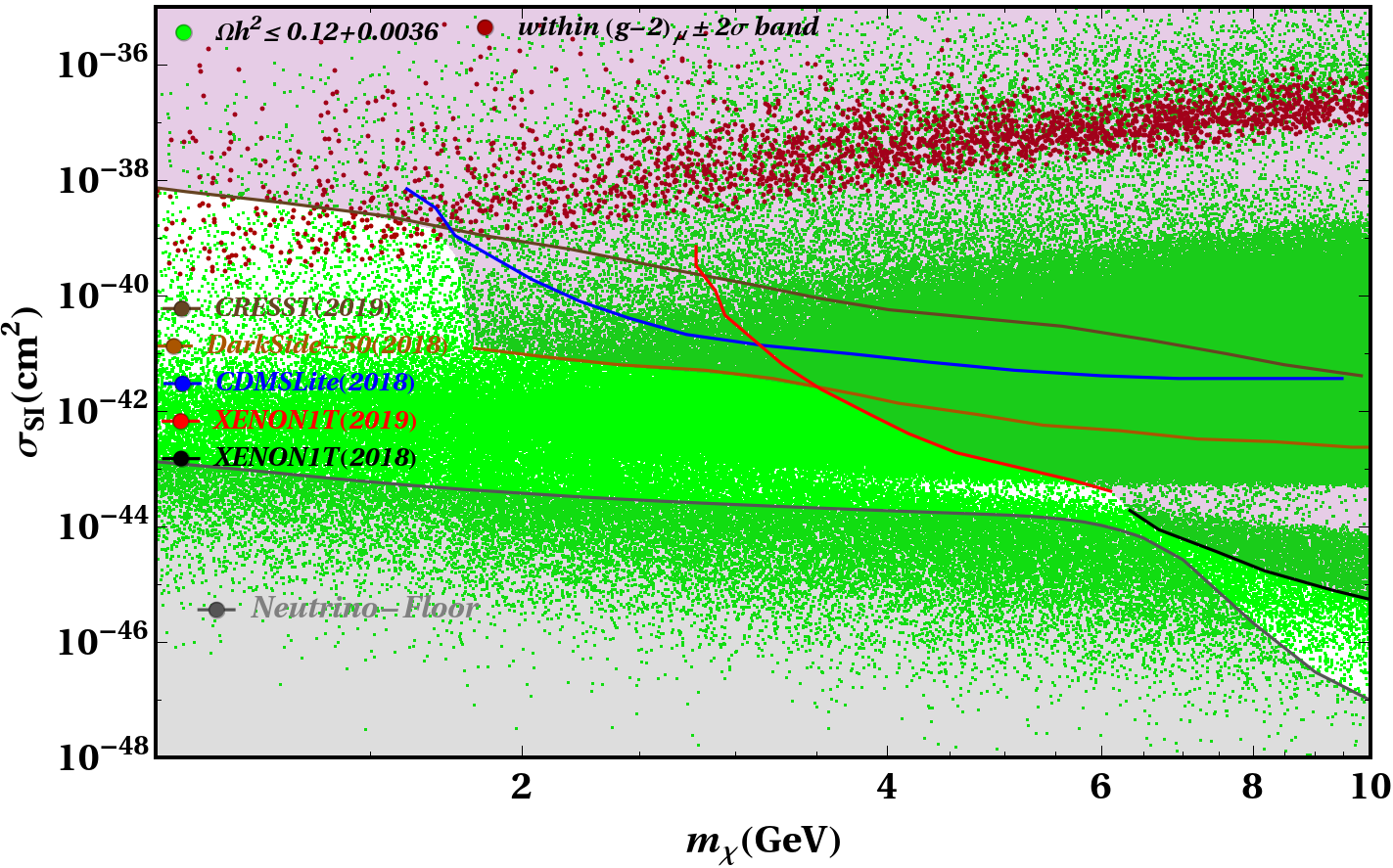}
\caption{\textit{(top) $\sigma_{SI}$ cross-section for the DM of mass 0.1 to 1 GeV, with other parameters as mentioned in the text. (bottom) Same DM-N cross-section for relatively higher DM masses from 1 to 10 GeV.}}
\label{DM_DD}
\end{figure}
\end{center}

In Fig(s) \ref{DM_DD-without (g-2)} - \ref{DM_DD}, we show direct-detection cross section ($\sigma_{SI}$) versus DM mass plot by considering certain benchmark values for the other parameters of the model. 
We analyze different DM mass regions without and with the inclusion of $(g-2)_\mu$ constraint to quantify its impact on the parameter space of the model. 
The purple-shaded regions depict the parameter space constrained by direct detection experiments. 
The primary constraint for sub-GeV DM is provided by the CRESST~\cite{CRESST:2019jnq} experiment. 
For DM in the range 1-10 GeV, the constraints become severe as we go towards higher mass, with the most constraining limits coming from Darkside-50~\cite{DarkSide:2018bpj}, XENON1T-2018~\cite{XENON:2018voc}, XENON1T-2019~\cite{XENON:2019rxp} along with SuperCDMS~\cite{SuperCDMS:2018gro}. 
The neutrino floor is depicted by the grey/cyan shaded region in both plots, with it becoming less constraining as we move towards high DM masses. 
For the points that go below the neutrino floor, the parameter space associated with them cannot be probed in this set of experiments, while those points are not necessarily ruled out. 
\vspace{-0.4cm}
\subsection{Without $(g-2)_\mu$ constraint}
\vspace{-0.4cm}
\begin{center}
\begin{figure}[!]
\includegraphics[scale=0.185]{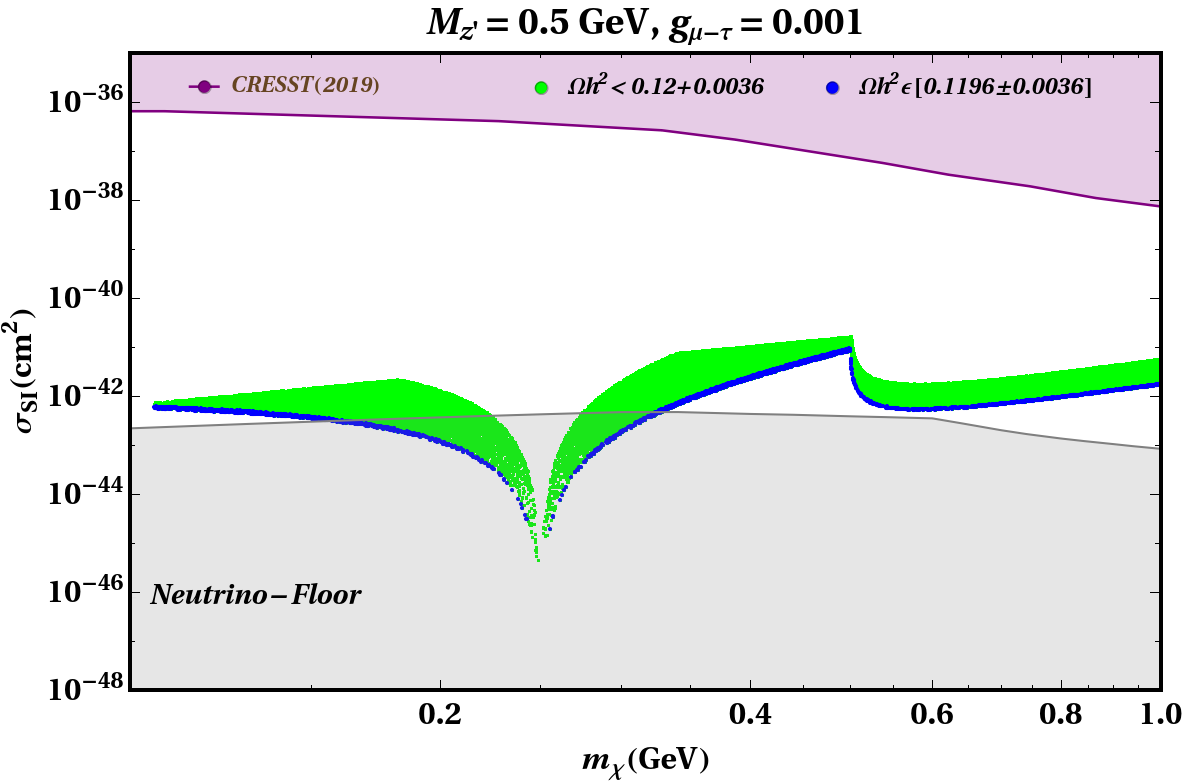}\hspace{0.2cm}
\includegraphics[scale=0.185]{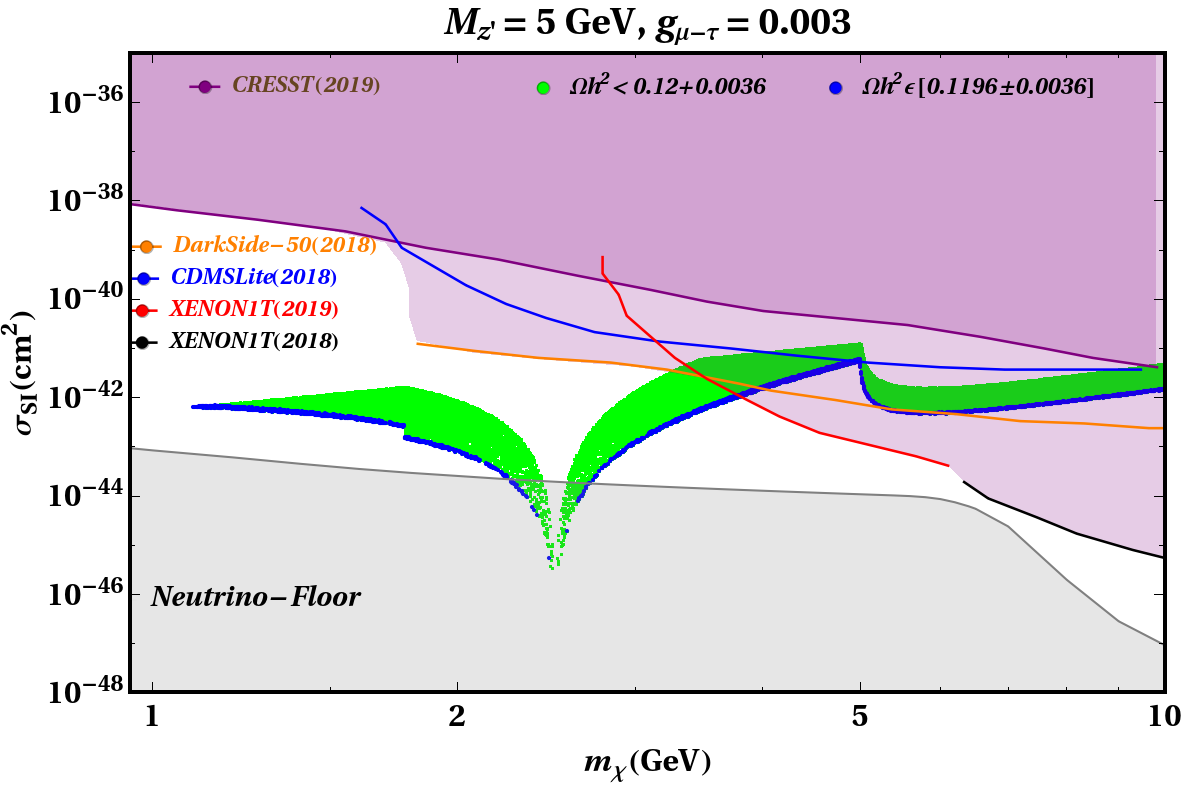}
\caption{\textit{ Before imposing $(g-2)_\mu$ constraint, (top) $\sigma_{SI}$ cross-section for the DM of mass 0.1 to 1 GeV, with parameters as mentioned in the text. (bottom) Same DM-N cross-section for relatively higher DM masses from 1 to 10 GeV.}}
\label{DM_DD-without (g-2)}
\end{figure}
\end{center}
\vspace{-0.4cm}
Without the imposition of the $(g-2)_\mu$ constraint, there is a lot of freedom to choose from the unshaded parameter space shown in Fig.~\ref{muon-g2}. 
We pick up two particular benchmark points in the ($M_{z'}$, $g_{\mu - \tau}$) parameter space to emphasize this point. 
For the low DM mass range of $0.1-1$~GeV, we choose a $M_{z'}$ mass of 0.5 GeV and $g_{\mu - \tau}$ coupling of $10^{-3}$. 
The upper plot in Fig.~\ref{DM_DD-without (g-2)} shows this particular benchmark scenario, where both s-channel and t-channel processes contribute to satisfy the correct relic density. 
The green points correspond to those parameter points which can satisfy at least $10 \%$ of the relic density whereas the blue points correspond to those satisfying the entire relic density. 
The dip at $m_\chi$ of 0.25 GeV corresponds to the s-channel resonance peak in the cross-section. 
The t-channel process opens around a DM mass of 0.5 GeV. 
We find a lot of parameter points that satisfy the relic density without being constrained by the direct-detection bounds.  

A similar analysis holds for the higher DM mass of 1-10 GeV mass as well. This is depicted in the lower plot of Fig.~\ref{DM_DD-without (g-2)} 
for a $M_{z'}$ mass of 5 GeV and $g_{\mu - \tau}$ coupling of $3 \times 10^{-3}$. 
A majority of points corresponding to the s-channel annihilation satisfy the relic density constraints in the allowed parameter space. 
The t-channel annihilation however falls foul of the direct-detection constraints. 
If we increase the DM mass, more and more points allowed by relic density (both s and t annihilation) get ruled out by the direct-detection constraints.

It is thus clear that before the imposition of the $(g-2)_\mu$ constraint, we have a lot of freedom in the sub-GeV DM mass range where both s and t channel annihilations are relevant for obtaining the correct relic density. 
This situation changes with the imposition of the $(g-2)_\mu$ constraint which will be elaborated in the subsequent part of this section.
\vspace{-0.4cm}
\subsection{With $(g-2)_\mu$ constraint: low $Z'$ mass}
\vspace{-0.4cm}
\begin{center}
\begin{figure}[!htb]
\includegraphics[scale=0.185]{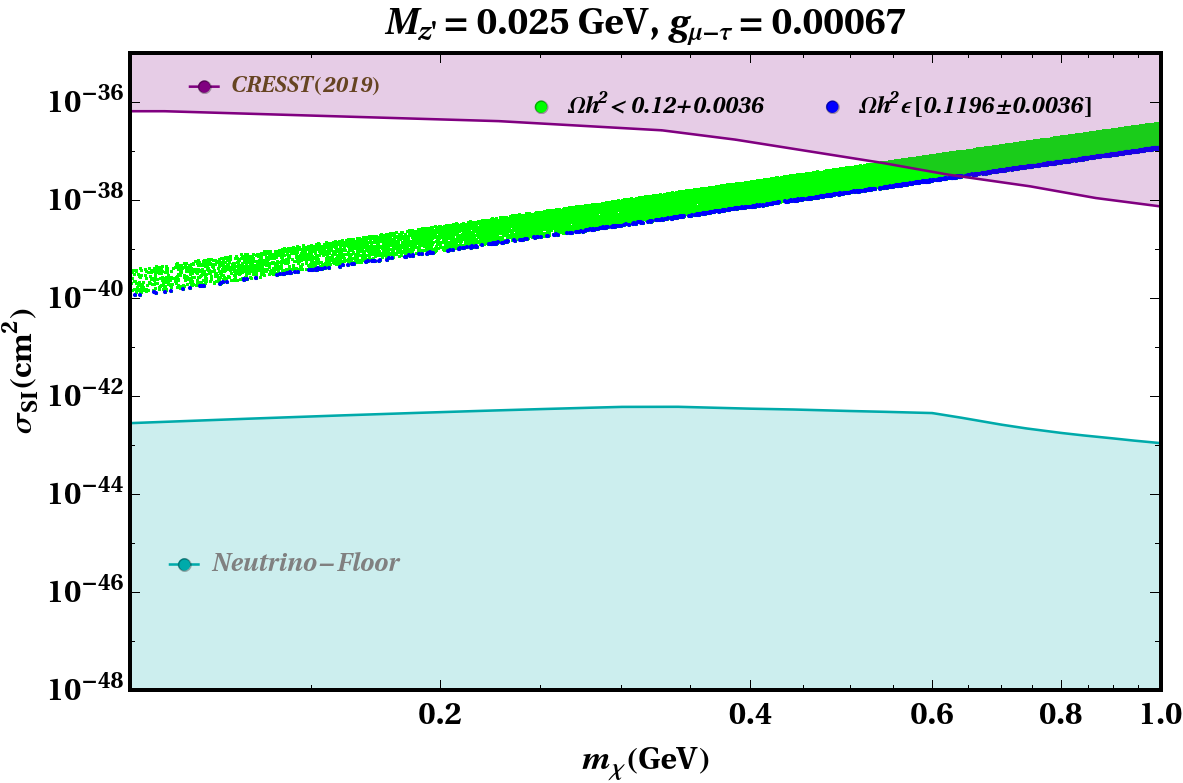}\hspace{0.2cm}
\includegraphics[scale=0.185]{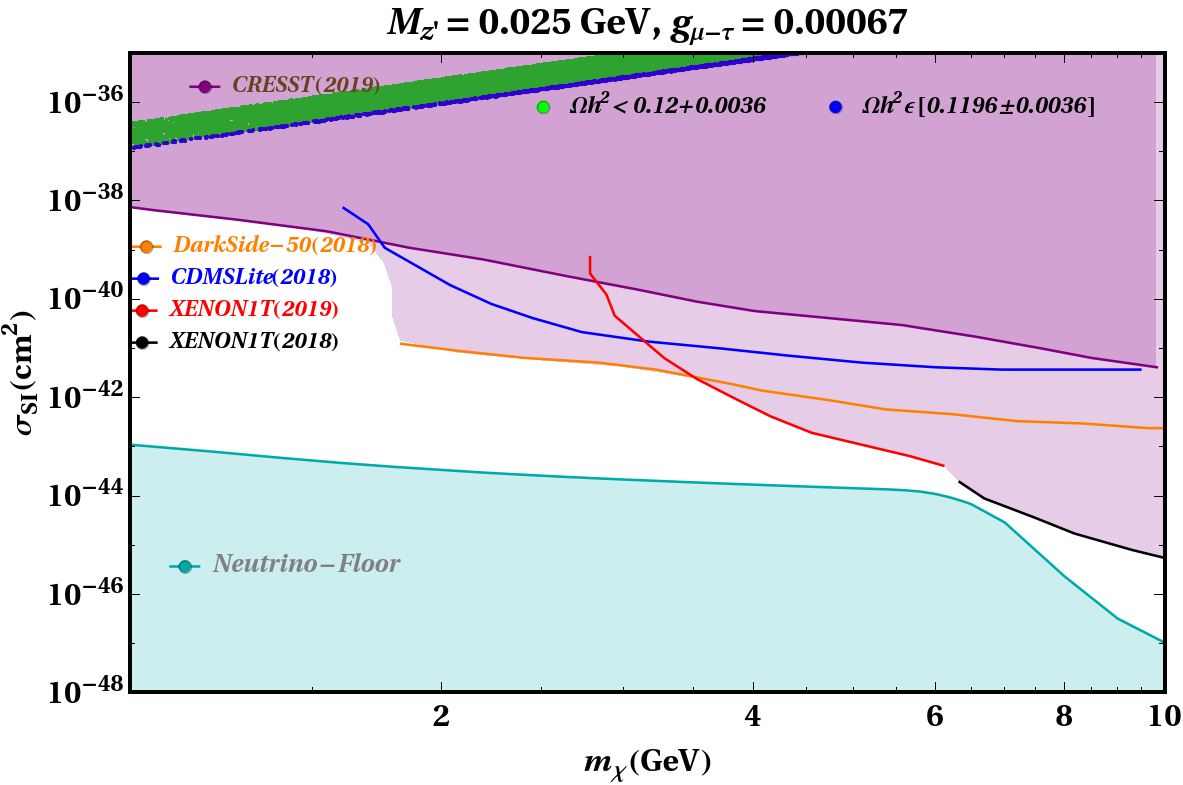}
\caption{\textit{(top) $\sigma_{SI}$ cross-section for the DM of mass 0.1 to 1 GeV, with parameters as mentioned in the text. (bottom) Same DM-N cross-section for relatively higher DM masses from 1 to 10 GeV.}}
\label{DM_DD-scenario1}
\end{figure}
\end{center}
\vspace{-0.5cm}
This scenario corresponds to a $Z^{\prime}$ mass of 25 MeV and a gauge coupling $g_{\mu-\tau}$ of $6.7 \times 10^{-4}$. With these values, we observe a significant enhancement of the neutrino floor which has already been discussed in detail in Section-\ref{nufloor}. The neutrino floor including its possible enhancement in this scenario is represented by the cyan-colored region in Figure~\ref{DM_DD-scenario1}. 
In the DM mass range of $0.1-1$~GeV, relic density can be satisfied very easily through the t-channel annihilation without violating any of the direct-detection constraints. 
The situation however becomes more constraining for a DM mass in the range of $1-10$~GeV, where the parameter region satisfying the thermal relic is entirely ruled out by the direct-detection experiments. 
This however does not rule out this parameter space in this model since with these small gauge couplings we still have the option for a freeze-in DM, which again we postpone for future dedicated work. 
More advanced DM direct detection experiments in the future with very low threshold recoil energies can become sensitive to lower direct-detection cross sections and hence can probe into the hitherto unexplored region of GeV scale DM. 
With the neutrino floor enhanced compared to the SM in this case, such detectors would also eventually start probing the neutrinos coming from this BSM scenario. 
However, in order to distinguish them from DM events, we either require detectors with angular sensitivity of the events or indirect detection methods. 
\vspace{-0.4cm}
 %can also help to distinguish the DM from the background neutrino events in the model parameter space with an enhanced neutrino floor.
\subsection{With $(g-2)_\mu$ constraint: high $Z'$ mass}
\vspace{-0.4cm}
\begin{center}
\begin{figure}[!htb]
\includegraphics[scale=0.185]{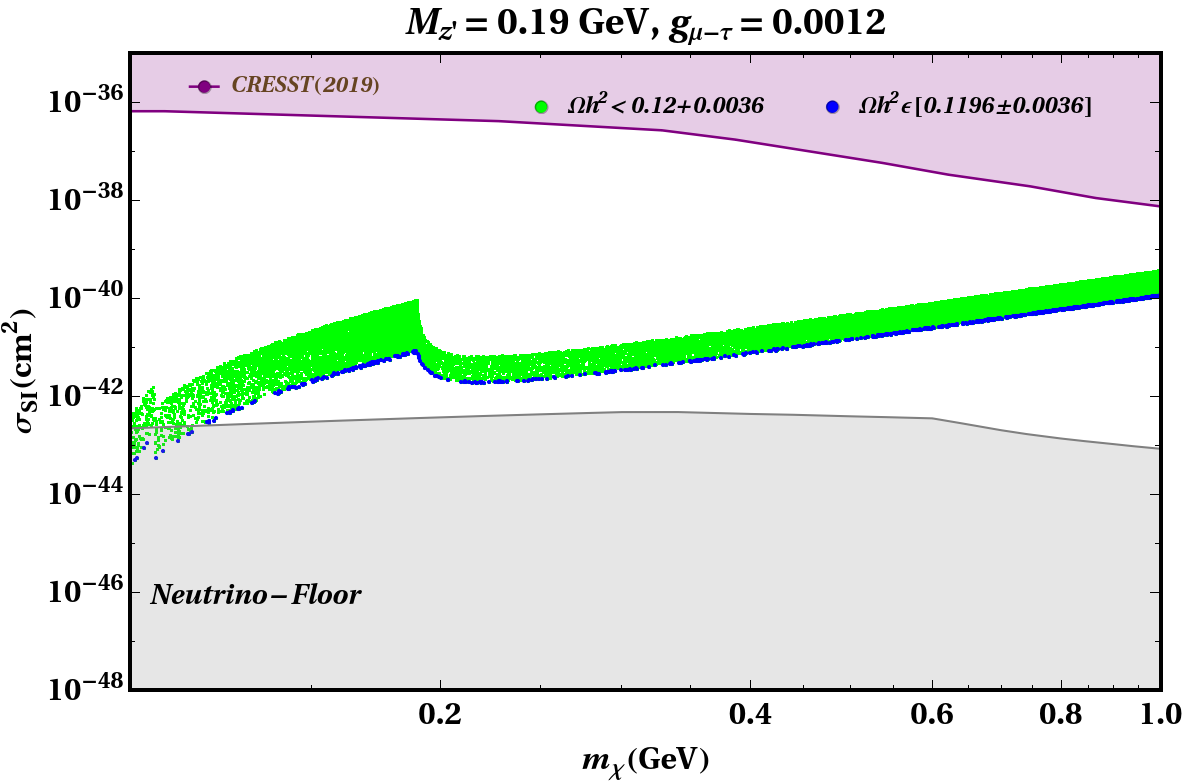}\hspace{0.2cm}
\includegraphics[scale=0.185]{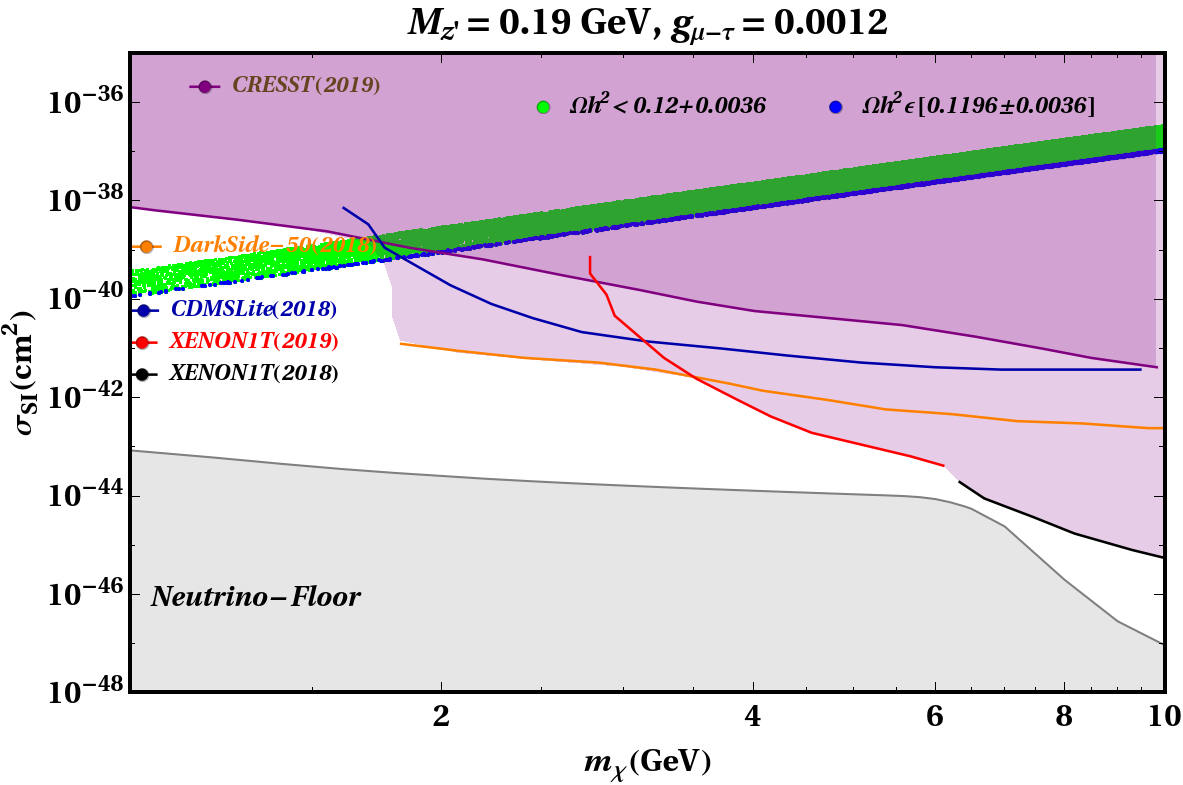}
\caption{\textit{(top) $\sigma_{SI}$ cross-section for the DM of mass 0.1 to 1 GeV, with parameters as mentioned in the text. (bottom) Same DM-N cross-section for relatively higher DM masses from 1 to 10 GeV.}}
\label{DM_DD-scenario2}
\end{figure}
\end{center}
\vspace{-0.4cm}
As already explained in Section-\ref{nufloor}, higher $Z'$ mass values decrease the enhancement in the neutrino floor. 
Although the BSM impact on the neutrino floor decreases, the prospect of a VLF DM detection brightens as we move towards higher $Z'$ mass. 
With heavier $Z'$, the direct-detection nucleon-DM scattering cross-section decreases and hence becomes less constrained from the experiments. 
This can be seen from the two plots of Figure-\ref{DM_DD-scenario2} for a benchmark point obtained from the rightmost allowed part of the muon $(g-2)$ band: $M_{z'}$ mass of 190 MeV and a gauge coupling $g_{\mu-\tau}$ of $1.2 \times 10^{-3}$.
Here the green and blue points span the parameter space in regions that arrange for under-abundant and exact DM relic density respectively.
In the DM mass range of $0.1-1$~GeV, till about 0.19 GeV, the tail of the s-channel resonant annihilation is efficient enough to satisfy the relic density constraints, after which the t-channel DM annihilation takes over.
In the DM mass range of 1-10 GeV, direct detection constraints allow only up to around 2 GeV DM mass.
As we will explain in the next section, the indirect detection bounds are much weaker than that of the muon $(g-2)$ in combination with the imposed direct-detection constraints. 
Hence, this particular region of parameter space is very important for future DM direct-detection studies. 
The neutrino floor is however not very interesting for a heavier $Z^{\prime}$ in as the modification of the floor is minimal here.
\vspace{-0.4cm}
\subsection{General parameter scan}
\vspace{-0.4cm}
\vspace{-0.4cm}
To decipher the full allowed parameter space, we look for direct detection cross-section numbers for the $M_{z'} \sim 0.01-2$~GeV, $g_{\mu-\tau} \sim 5\times 10^{-4} - 5\times 10^{-3}$ and the DM gauge coupling $g_\chi$ from $5 \times 10^{-4}$ to $2.5 \times 10^{-1}$. 
Since the neutrino floor changes for each value of $M_{z'}$ and $g_{\mu-\tau}$, we depict only the SM neutrino floor, as neutrino floor modification effects were already shown earlier. 
Here the green points denote the cross sections satisfying at least $10\%$ of the DM relic density before the imposition of the $(g-2)_{\mu}$ constraint. 
The dark-red points correspond to the subset of the green points which survive after imposing the $\pm 2 \sigma$ $(g-2)_\mu$ bound.

Many parameter points can satisfy the imposed relic density constraint both in the sub-GeV and super-GeV DM mass range while simultaneously steering clear of the direct-detection constraints and the neutrino floor. 
The $(g-2)_\mu$ bound, however, mostly prefers only the sub-GeV DM mass range, and super-GeV DM gets heavily constrained from the DM direct-detection searches.

\section{Constraints from Indirect Detection}
\label{indirect}
The indirect detection constraints in general can provide strong bounds for the dark matter parameter space~\cite{Leane:2018kjk}. 
Some of the most stringent bounds come from the Planck measurement of CMB anisotropies~\cite{Planck:2015fie}, where injection of energy through ionizing charged particles from the DM annihilation can perturb the standard thermal history of ionization involving the hydrogen and helium gas as the constituents of the Universe at that era. 
Gamma-ray fluxes from Dwarf Spheroidal Galaxies (dSphs) which are measured by Fermi-Lat~\cite{Albert_2017} can also provide strong bounds to any photonic final states from DM annihilation. 
Along with these, the AMS measurements~\cite{PhysRevLett.110.141102} of the electron and positron flux can also provide strong constraints for electron and other leptonic final states. 
Due to the inefficient annihilation in the s-channel to the muons and taus in our case, constraints coming from bounds on annihilation cross-section to a pair of muons and taus are insignificant for our analysis. 
The constraints for annihilation to two electrons or positrons are even weaker due to the absence of any tree-level coupling of $Z'$ to electrons. 
Due to the presence of t-channel annihilations to a pair of $Z'$s, the strongest limits emanate from the dSphs measurement of $4 \tau$ final states once DM mass goes above 3 GeV~\cite{Profumo:2017obk}. 
Compared to this, the strongest bound from the $4\mu$ final states coming from the CMB measurements are slightly weaker and come into play once we are above 4 GeV DM mass~\cite{Profumo:2017obk}. 
These bounds, although strong, cannot compete with the bounds coming from the combination of $(g-2)_\mu$ and direct detection measurements as shown in Figure~\ref{DM_DD-scenario1} and Figure~\ref{DM_DD-scenario2}. 
Hence the constraints coming from the anomalous magnetic moment of the muon and dark matter direct detection constraints provide complementarity among themselves and constrain the DM sector more stringently than the constraints coming from the indirect detection experiments.

\section{Summary and Conclusion}
\label{summary}
In this article, we reinvestigate the viability of $U(1)_{L_{\mu}-L_{\tau}}$ model to simultaneously explain recent anomalous $(g-2)_\mu$ and DM relic density, along with possible modification in the neutrino floor. 
The feasible parameter range of $M_{z'}$ in $10-200$~MeV with $g_{\mu-\tau} \sim (0.5-1.0)\times 10^{-3}$, can explain anomalous $(g-2)_\mu$ measured by the most recent FNAL measurement.
This parameter space is also found to be consistent with the earlier anomalous result from the FNAL. 
We explore this allowed region for possible modification of the neutrino floor to find out that with increasing $M_{z'}$ along the $(g-2)_{\mu}$ viable parameter region, neutrino floor enhancement becomes less significant. 
The neutrino floor enhancement compared to the SM appears to be in the range $1.7-2.7$ in the absence of anomalous $(g-2)_\mu$ for Germanium and Xenon-based experiments. 
With the inclusion of $(g-2)_\mu$ constraint, the new allowed benchmark point $\left(M_{z'}=16\, \rm MeV, g_{\mu-\tau}=6.0\times 10^{-4}\right)$ shows the maximum enhancement of $1.6-2.2$ times for different DM direct detection experiments. 
Next, considering the ultralight DM bound by cosmology, we venture into a territory exploring the GeV scale dark matter. 
We show how the constraints from $(g-2)_\mu$, affect DM phenomenology of a vectorlike fermion (VLF) DM of masses from 0.1 to 10 GeV.
The DM candidate $\chi$ contributes to dark matter relic density though the s-channel annihilation process $\bar{\chi} \,\chi \rightarrow Z'\rightarrow \bar{l}\,\,l\,(\bar{\nu}_l\,\, \nu) $ and t-channel process  $\bar{\chi} \,\chi \rightarrow Z'\,\,Z' $ with t-channel playing the dominant role in almost entire part of the parameter space. 
For sub-GeV DM with $m_\chi$: $\left( 0.1-1\right)$ GeV in the absence of the anomalous $(g-2)_\mu$, the DM relic density allows for smaller coupling values and mass in the range of $M_{z'}=0.01-10$ GeV. 
With the inclusion of $(g-2)_\mu$ constraint allowed parameter space is more restricted with  $m_{Z'}=0.01-0.2$ GeV and $g_\chi$ preferring relatively higher values, when we explore the sub-GeV DM mass range.
We found a similar trend for the super GeV range, but with the inclusion of $(g-2)_\mu$, parameter space is even more restricted, with $g_\chi$ being confined to a narrow band of values. 
For sub-GeV dark matter, the CRESST experiment excludes $\sigma_{SI}$ above $10^{-37}-10^{-38}$ cm$^2$, but still appreciable parameter space in this range is allowed. 
In order to elaborate things more, we include two benchmark scenarios: one towards the leftmost part of the allowed $(g-2)_\mu$ band and the other towards the rightmost part, along with the behavior of the corresponding neutrino floor in those regions. 
We observe from the DM parameter space constrained by direct detection experiments, that most of the super-GeV parameter space viable for $(g-2)_\mu$ is ruled out by DarKSide-50, CDMSLite, XENON-1T, with only small parameter space being allowed for DM mass less than 2 GeV. 
One important observation is that the constraints are very different in the different parts of the muon $(g-2)$ allowed region.

In conclusion, the model parameter space viable for $g-2$ experiments can only account for a tiny part of sub-GeV VLF dark matter, while showing around two times enhancement of the neutrino floor as well.
 This region of interest can be worthwhile to probe in upcoming neutrino experiments like COHERENT. 
  With increasing exposure, direct detection experiments may find it difficult to differentiate between the dark matter signal from the neutrino background. 
  The conclusions we draw here are valid for the Dirac fermionic DM candidates. 
  A light Majorana fermion DM candidate can have a new scalar portal for its annihilation, which can potentially open up most of the parameter space we restrict due to the muon $(g-2)$ constraint here.
Also a scalar dark matter can be studied in this context. 
In general GeV scale DM is an interesting region due to its viability, being not so constrained by other existing constraints. 
The recent anomalous $(g-2)$ measurement plays a crucial role when coupled with the DM direct detection constraints, more constraining to the GeV scale DM than the usual indirect detection constraints. 
\vspace{-0.5cm} 
%%%%%%%%%%%%%%%%  
\section*{Acknowledgments}
\vspace{-0.3cm}
 We would like to thank Prof. Debajyoti Choudhury for the useful discussions.  K.D. acknowledges the Council for Scientific and Industrial Research
(CSIR), India for JRF/SRF fellowship with award letter
no. 09/045(1654)/2019-EMR-I.
SS thanks Vivekananda Centre for Research (VCR) for providing the research facilities. 
MPS would like to thank S.P. Singh for partial financial support.
\vspace{-0.8cm}
\appendix
\section*{Appendix}
\vspace{-0.1cm}
\subsection*{Enhancement in CE$\nu$NS rate and neutrino floor}
\label{tables}
I. {\bf Benchmark
chosen: $Z'$   mass 16 MeV and coupling  $g_{\mu-\tau}= 6.0\times10^{-4}$} 
%\caption{•}
\begin{table}[h!]
\centering
\begin{tabular}{ |c|c|c|c|c| } 
\hline
m$_{DM}$(GeV) & SM N. floor(cm$^{2}$) &  U(1)$_{\mu-\tau}$ N. floor(cm$^{2}$) & Enh. \\
\hline
%\centering
\multicolumn{4}{| c |}{\bf Germanium based experiments}\\
\hline
0.5 & 2.36$\times 10^{-43}$ & 5.22$\times 10^{-43}$  & 2.211  \\
\hline 
1. &  5.23$\times 10^{-44}$ & 1.14$\times 10^{-43}$  & 2.17  \\ 
\hline
5 & 7.05$\times 10^{-45}$ & 1.51$\times 10^{-44}$  & 2.121  \\ 
\hline
10 & 1.42$\times 10^{-47}$ & 1.74$\times 10^{-47}$  & 1.225 \\

\hline 
\multicolumn{4}{| c |}{\bf Xenon based experiments}\\
\hline
0.5 & 3.91$\times 10^{-43}$ & 6.4$\times 10^{-43}$  & 1.63  \\
\hline 
1. &  8.49$\times 10^{-44}$ & 1.38$\times 10^{-43}$  & 1.62  \\ 
\hline
5 & 1.08$\times 10^{-44}$ & 1.76$\times 10^{-44}$  & 1.62  \\ 
\hline
10 & 9.27$\times 10^{-48}$ & 1.04$\times 10^{-47}$  & 1.12 \\
\hline
\end{tabular}
\caption{ \small \em{Neutrino floor versus dark matter mass table highlighting modification of neutrino floor for U$_{\mu-\tau}$  with respect to SM. We have used the abbreviations  N. floor and Enh.  for Neutrino floor and Enhancement respectively.  } }\label{table:4}
\end{table}
%\vspace{5cm}

II. {\bf Benchmark chosen: $Z'$   mass 200 MeV and coupling  $g_{\mu-\tau}= 1.64\times10^{-3}$}
\begin{table}[h!]
\centering
\begin{tabular}{ |c|c|c|c|c| } 
\hline
m$_{DM}$(GeV) & SM N. floor(cm$^{2}$) &  U(1)$_{\mu-\tau}$ N. floor(cm$^{2}$) & Enh. \\
\hline 
\multicolumn{4}{| c |}{\bf Germanium based experiments}\\
\hline
0.5 & 2.36$\times 10^{-43}$ & 2.46$\times 10^{-43}$  & 1.04  \\
\hline 
1. &  5.23$\times 10^{-44}$ & 5.48$\times 10^{-43}$  & 1.047  \\ 
\hline
5 & 7.05$\times 10^{-45}$ & 7.44$\times 10^{-44}$  & 1.055  \\ 
\hline
10 & 1.42$\times 10^{-47}$ & 1.49$\times 10^{-47}$  & 1.049 \\

\hline 
\multicolumn{4}{| c |}{\bf Xenon based experiments}\\
\hline
0.5 & 3.91$\times 10^{-43}$ & 4.01$\times 10^{-43}$  & 1.02  \\
\hline 
1. &  8.49$\times 10^{-44}$ & 8.74$\times 10^{-43}$  & 1.03  \\ 
\hline
5 & 1.08$\times 10^{-44}$ & 1.125$\times 10^{-44}$  & 1.04  \\ 
\hline
10 & 9.27$\times 10^{-48}$ & 9.56$\times 10^{-47}$  & 1.03 \\
\hline
\end{tabular}
\caption{ \small \em{Neutrino floor versus dark matter mass table highlighting modification of neutrino floor for U$_{\mu-\tau}$  with respect to SM. We have used the abbreviations  N. floor and Enh.  for Neutrino floor and Enhancement respectively.} }\label{table:4}
\end{table}

\pagebreak

III. {\bf Benchmark
chosen: $Z'$  mass 50 MeV and coupling  $g_{\mu-\tau}= 1.1\times10^{-3}$} 
\begin{table}[h!]
\centering
\begin{tabular}{ |c|c|c|c|c| } 
\hline
m$_{DM}$(GeV) & SM N. floor(cm$^{2}$) &  U(1)$_{\mu-\tau}$ N. floor(cm$^{2}$) & Enh. \\
\hline
\multicolumn{4}{| c |}{\bf Germanium based experiments}\\
\hline
0.5 & 2.36$\times 10^{-43}$ & 3.16$\times 10^{-43}$  & 1.33 \\
\hline 
1. &  5.23$\times 10^{-44}$ & 7.23$\times 10^{-43}$  & 1.38 \\ 
\hline
5 & 7.05$\times 10^{-45}$ & 1.01$\times 10^{-44}$  & 1.43  \\ 
\hline
10 & 1.42$\times 10^{-47}$ & 1.82$\times 10^{-47}$  & 1.28 \\
\hline 

\multicolumn{4}{| c |}{\bf Xenon based experiments}\\
\hline

0.5 & 3.91$\times 10^{-43}$ & 4.67$\times 10^{-43}$  & 1.19  \\
\hline 
1. &  8.49$\times 10^{-44}$ & 1.03$\times 10^{-43}$  & 1.21  \\ 
\hline
5 & 1.08$\times 10^{-44}$ & 1.35$\times 10^{-44}$  & 1.25  \\ 
\hline
10 & 9.27$\times 10^{-48}$ & 1.08$\times 10^{-47}$  & 1.16 \\
\hline 
\end{tabular}
\caption{ \small \em{Neutrino floor versus dark matter mass table highlighting modification of neutrino floor for U$_{\mu-\tau}$  with respect to SM. We have used the abbreviations  N. floor and Enh.  for Neutrino floor and Enhancement respectively. } }\label{table:4}
\end{table}
%\newpage
IV. {\bf Benchmark chosen: $Z'$  mass 25 MeV and coupling  $g_{\mu-\tau}= 0.93\times10^{-3}$}
\begin{table}[h!]
\centering
\begin{tabular}{ |c|c|c|c|c| } 
\hline
m$_{DM}$(GeV) & SM N. floor(cm$^{2}$) &  U(1)$_{\mu-\tau}$ N. floor(cm$^{2}$) & Enh. \\
\hline
\multicolumn{4}{| c |}{\bf Germanium based experiments}\\

\hline
0.5 & 2.36$\times 10^{-43}$ & 5.06$\times 10^{-43}$  & 2.14  \\
\hline 
1. &  5.23$\times 10^{-44}$ & 1.16$\times 10^{-43}$  & 2.23  \\ 
\hline
5 & 7.05$\times 10^{-45}$ & 1.61$\times 10^{-44}$  & 2.28  \\ 
\hline
10 & 1.42$\times 10^{-47}$ & 2.05$\times 10^{-47}$  & 1.443 \\
\hline 
\multicolumn{4}{| c |}{\bf Xenon based experiments}\\
\hline
0.5 & 3.91$\times 10^{-43}$ & 6.27$\times 10^{-43}$  & 1.60  \\
\hline 
1. &  8.49$\times 10^{-44}$ & 1.40 $\times 10^{-43}$  & 1.65  \\ 
\hline
5 & 1.08$\times 10^{-44}$ & 1.84$\times 10^{-44}$  & 1.70  \\ 
\hline
10 & 9.27$\times 10^{-48}$ & 1.16$\times 10^{-47}$  & 1.25 \\
\hline 
\end{tabular}
\caption{ \small \em{Neutrino floor versus dark matter mass table highlighting modification of neutrino floor for U$_{\mu-\tau}$  with respect to SM. We have used the abbreviations  N. floor and Enh.  for Neutrino floor and Enhancement respectively. } }\label{table:4}
\end{table}

\newpage

\bibliographystyle{unsrtnat}

\bibliography{ref}

\begin{thebibliography}{52}
\providecommand{\natexlab}[1]{#1}
\providecommand{\url}[1]{\texttt{#1}}
\expandafter\ifx\csname urlstyle\endcsname\relax
  \providecommand{\doi}[1]{doi: #1}\else
  \providecommand{\doi}{doi: \begingroup \urlstyle{rm}\Url}\fi

\bibitem[Aguillard et~al.(2023)]{Muong-2:2023cdq}
D.~P. Aguillard et~al.
\newblock {Measurement of the Positive Muon Anomalous Magnetic Moment to 0.20
  ppm}.
\newblock 8 2023.
\newblock \doi{https://arxiv.org/pdf/2308.06230.pdf}.

\bibitem[Abi et~al.(2021)]{Muong-2:2021ojo}
B.~Abi et~al.
\newblock {Measurement of the Positive Muon Anomalous Magnetic Moment to 0.46
  ppm}.
\newblock \emph{Phys. Rev. Lett.}, 126\penalty0 (14):\penalty0 141801, 2021.
\newblock \doi{10.1103/PhysRevLett.126.141801}.

\bibitem[Girotti(2022)]{Girotti:2021vyl}
Paolo Girotti.
\newblock {Status of the Fermilab Muon g \textendash{} 2 Experiment}.
\newblock \emph{EPJ Web Conf.}, 262:\penalty0 01003, 2022.
\newblock \doi{10.1051/epjconf/202226201003}.

\bibitem[Bennett et~al.(2006)]{Muong-2:2006rrc}
G.~W. Bennett et~al.
\newblock {Final Report of the Muon E821 Anomalous Magnetic Moment Measurement
  at BNL}.
\newblock \emph{Phys. Rev. D}, 73:\penalty0 072003, 2006.
\newblock \doi{10.1103/PhysRevD.73.072003}.

\bibitem[Bauer et~al.(2018)Bauer, Foldenauer, and Jaeckel]{Bauer:2018onh}
Martin Bauer, Patrick Foldenauer, and Joerg Jaeckel.
\newblock {Hunting All the Hidden Photons}.
\newblock \emph{JHEP}, 07:\penalty0 094, 2018.
\newblock \doi{10.1007/JHEP07(2018)094}.
\newblock [JHEP18,094(2020)].

\bibitem[Qi et~al.(2021)Qi, Yang, Liu, and Sun]{Qi:2021rhh}
XinXin Qi, AiGeng Yang, Wei Liu, and Hao Sun.
\newblock {Scalar dark matter and Muon g-2 in a $U(1)_{L_{\mu}-L_{\tau}}$
  model}.
\newblock 6 2021.

\bibitem[Hapitas et~al.(2022)Hapitas, Tuckler, and Zhang]{Hapitas:2021ilr}
Timothy Hapitas, Douglas Tuckler, and Yue Zhang.
\newblock {General kinetic mixing in gauged
  U(1)L\ensuremath{\mu}-L\ensuremath{\tau} model for muon g-2 and dark matter}.
\newblock \emph{Phys. Rev. D}, 105\penalty0 (1):\penalty0 016014, 2022.
\newblock \doi{10.1103/PhysRevD.105.016014}.

\bibitem[Gninenko and Krasnikov(2018)]{Gninenko:2018tlp}
S.~N. Gninenko and N.~V. Krasnikov.
\newblock {Probing the muon $g_\mu$ - 2 anomaly, $L_\mu - L_\tau$ gauge boson
  and Dark Matter in dark photon experiments}.
\newblock \emph{Phys. Lett. B}, 783:\penalty0 24--28, 2018.
\newblock \doi{10.1016/j.physletb.2018.06.043}.

\bibitem[Borah et~al.(2021)Borah, Dutta, Mahapatra, and Sahu]{Borah:2021jzu}
Debasish Borah, Manoranjan Dutta, Satyabrata Mahapatra, and Narendra Sahu.
\newblock {Muon (g \ensuremath{-} 2) and XENON1T excess with boosted dark
  matter in L\ensuremath{\mu} \ensuremath{-} L\ensuremath{\tau} model}.
\newblock \emph{Phys. Lett. B}, 820:\penalty0 136577, 2021.
\newblock \doi{10.1016/j.physletb.2021.136577}.

\bibitem[Borah et~al.(2022)Borah, Dutta, Mahapatra, and Sahu]{Borah:2021khc}
Debasish Borah, Manoranjan Dutta, Satyabrata Mahapatra, and Narendra Sahu.
\newblock {Lepton anomalous magnetic moment with singlet-doublet fermion dark
  matter in a scotogenic U(1)L\ensuremath{\mu}-L\ensuremath{\tau} model}.
\newblock \emph{Phys. Rev. D}, 105\penalty0 (1):\penalty0 015029, 2022.
\newblock \doi{10.1103/PhysRevD.105.015029}.

\bibitem[Ko et~al.(2021)Ko, Nomura, and Okada]{Ko:2021lpx}
P.~Ko, Takaaki Nomura, and Hiroshi Okada.
\newblock {Muon $g-2$, $B\to K^{(*)}\mu^+ \mu^-$ anomalies, and leptophilic
  dark matter in $U(1)_{\mu-\tau}$ gauge symmetry}.
\newblock 10 2021.

\bibitem[Holst et~al.(2021)Holst, Hooper, and Krnjaic]{Holst:2021lzm}
Ian Holst, Dan Hooper, and Gordan Krnjaic.
\newblock {The Simplest and Most Predictive Model of Muon $g-2$ and Thermal
  Dark Matter}.
\newblock 7 2021.

\bibitem[Baek(2016)]{Baek:2015fea}
Seungwon Baek.
\newblock {Dark matter and muon $(g-2)$ in local $U(1)_{L_\mu-L_\tau}$-extended
  Ma Model}.
\newblock \emph{Phys. Lett.}, B756:\penalty0 1--5, 2016.
\newblock \doi{10.1016/j.physletb.2016.02.062}.

\bibitem[Patra et~al.(2017)Patra, Rao, Sahoo, and Sahu]{Patra:2016shz}
Sudhanwa Patra, Soumya Rao, Nirakar Sahoo, and Narendra Sahu.
\newblock {Gauged $U(1)_{L_\mu - L_\tau}$ model in light of muon $g-2$ anomaly,
  neutrino mass and dark matter phenomenology}.
\newblock \emph{Nucl. Phys.}, B917:\penalty0 317--336, 2017.
\newblock \doi{10.1016/j.nuclphysb.2017.02.010}.

\bibitem[Foldenauer(2019)]{Foldenauer:2018zrz}
Patrick Foldenauer.
\newblock Light dark matter in a gauged $u(1)_{L_\mu-L_\tau}$ model.
\newblock \emph{Phys. Rev.}, D99\penalty0 (3):\penalty0 035007, 2019.
\newblock \doi{10.1103/PhysRevD.99.035007}.

\bibitem[Costa et~al.(2022)Costa, Khan, and Kim]{Costa:2022oaa}
Francesco Costa, Sarif Khan, and Jinsu Kim.
\newblock {A Two-Component Dark Matter Model and its Associated Gravitational
  Waves}.
\newblock 2 2022.

\bibitem[Sadhukhan and Singh(2021)]{Sadhukhan:2020etu}
Soumya Sadhukhan and Manvinder~Pal Singh.
\newblock {Neutrino floor in leptophilic $U(1)$ models: Modification in
  $U(1)_{L_{\mu}-L_{\tau}}$}.
\newblock \emph{Phys. Rev. D}, 103\penalty0 (1):\penalty0 015015, 2021.
\newblock \doi{10.1103/PhysRevD.103.015015}.

\bibitem[Foot(1991)]{Foot:1990mn}
Robert Foot.
\newblock {New Physics From Electric Charge Quantization?}
\newblock \emph{Mod. Phys. Lett.}, A6:\penalty0 527--530, 1991.
\newblock \doi{10.1142/S0217732391000543}.

\bibitem[He et~al.(1991{\natexlab{a}})He, Joshi, Lew, and Volkas]{He:1990pn}
X.~G. He, Girish~C. Joshi, H.~Lew, and R.~R. Volkas.
\newblock {NEW Z-prime PHENOMENOLOGY}.
\newblock \emph{Phys. Rev.}, D43:\penalty0 22--24, 1991{\natexlab{a}}.
\newblock \doi{10.1103/PhysRevD.43.R22}.

\bibitem[He et~al.(1991{\natexlab{b}})He, Joshi, Lew, and Volkas]{He:1991qd}
Xiao-Gang He, Girish~C. Joshi, H.~Lew, and R.~R. Volkas.
\newblock {Simplest Z-prime model}.
\newblock \emph{Phys. Rev.}, D44:\penalty0 2118--2132, 1991{\natexlab{b}}.
\newblock \doi{10.1103/PhysRevD.44.2118}.

\bibitem[Choudhury et~al.(2020)Choudhury, Deka, Mandal, and
  Sadhukhan]{Choudhury:2020cpm}
Debajyoti Choudhury, Kuldeep Deka, Tanumoy Mandal, and Soumya Sadhukhan.
\newblock {Neutrino and $Z'$ phenomenology in an anomaly-free $\mathbf{U}(1)$
  extension: role of higher-dimensional operators}.
\newblock \emph{JHEP}, 06:\penalty0 111, 2020.
\newblock \doi{10.1007/JHEP06(2020)111}.

\bibitem[et.all{\'{a}}(2017)]{Albert_2017}
A.~Albert et.all{\'{a}}.
\newblock {SEARCHING} {FOR} {DARK} {MATTER} {ANNIHILATION} {IN} {RECENTLY}
  {DISCOVERED} {MILKY} {WAY} {SATELLITES} {WITHFERMI}-{LAT}.
\newblock \emph{The Astrophysical Journal}, 834\penalty0 (2):\penalty0 110, jan
  2017.
\newblock \doi{10.3847/1538-4357/834/2/110}.
\newblock URL \url{https://doi.org/10.3847/1538-4357/834/2/110}.

\bibitem[Chang et~al.(2017)]{TheDAMPE:2017dtc}
J.~Chang et~al.
\newblock {The DArk Matter Particle Explorer mission}.
\newblock \emph{Astropart. Phys.}, 95:\penalty0 6--24, 2017.
\newblock \doi{10.1016/j.astropartphys.2017.08.005}.

\bibitem[Aguilar(2013)]{PhysRevLett.110.141102}
M.~.et~all Aguilar.
\newblock First result from the alpha magnetic spectrometer on the
  international space station: Precision measurement of the positron fraction
  in primary cosmic rays of 0.5--350 gev.
\newblock \emph{Phys. Rev. Lett.}, 110:\penalty0 141102, Apr 2013.
\newblock \doi{10.1103/PhysRevLett.110.141102}.
\newblock URL \url{https://link.aps.org/doi/10.1103/PhysRevLett.110.141102}.

\bibitem[Chun et~al.(2019)Chun, Das, Kim, and Kim]{Chun:2018ibr}
Eung~Jin Chun, Arindam Das, Jinsu Kim, and Jongkuk Kim.
\newblock {Searching for flavored gauge bosons}.
\newblock \emph{JHEP}, 02:\penalty0 093, 2019.
\newblock \doi{10.1007/JHEP02(2019)093}.
\newblock [Erratum: JHEP 07, 024 (2019)].

\bibitem[Riordan et~al.(1987)]{Riordan:1987aw}
E.~M. Riordan et~al.
\newblock {A Search for Short Lived Axions in an Electron Beam Dump
  Experiment}.
\newblock \emph{Phys. Rev. Lett.}, 59:\penalty0 755, 1987.
\newblock \doi{10.1103/PhysRevLett.59.755}.

\bibitem[Bjorken et~al.(1988)Bjorken, Ecklund, Nelson, Abashian, Church, Lu,
  Mo, Nunamaker, and Rassmann]{Bjorken:1988as}
J.~D. Bjorken, S.~Ecklund, W.~R. Nelson, A.~Abashian, C.~Church, B.~Lu, L.~W.
  Mo, T.~A. Nunamaker, and P.~Rassmann.
\newblock {Search for Neutral Metastable Penetrating Particles Produced in the
  SLAC Beam Dump}.
\newblock \emph{Phys. Rev.}, D38:\penalty0 3375, 1988.
\newblock \doi{10.1103/PhysRevD.38.3375}.

\bibitem[Bross et~al.(1991)Bross, Crisler, Pordes, Volk, Errede, and
  Wrbanek]{Bross:1989mp}
A.~Bross, M.~Crisler, Stephen~H. Pordes, J.~Volk, S.~Errede, and J.~Wrbanek.
\newblock {A Search for Shortlived Particles Produced in an Electron Beam
  Dump}.
\newblock \emph{Phys. Rev. Lett.}, 67:\penalty0 2942--2945, 1991.
\newblock \doi{10.1103/PhysRevLett.67.2942}.

\bibitem[et~al(1986)]{DORENBOSCH1986473}
J.~Dorenbosch et~al.
\newblock A search for decays of heavy neutrinos in the mass range 0.5--2.8
  gev.
\newblock \emph{Physics Letters B}, 166\penalty0 (4):\penalty0 473--478, 1986.
\newblock ISSN 0370-2693.
\newblock \doi{https://doi.org/10.1016/0370-2693(86)91601-1}.
\newblock URL
  \url{https://www.sciencedirect.com/science/article/pii/0370269386916011}.

\bibitem[Athanassopoulos et~al.(1998)]{LSND:1997vqj}
C.~Athanassopoulos et~al.
\newblock {Evidence for muon-neutrino ---\ensuremath{>} electron-neutrino
  oscillations from pion decay in flight neutrinos}.
\newblock \emph{Phys. Rev. C}, 58:\penalty0 2489--2511, 1998.
\newblock \doi{10.1103/PhysRevC.58.2489}.

\bibitem[Blumlein and Brunner(2011)]{Blumlein:2011mv}
Johannes Blumlein and Jurgen Brunner.
\newblock {New Exclusion Limits for Dark Gauge Forces from Beam-Dump Data}.
\newblock \emph{Phys. Lett. B}, 701:\penalty0 155--159, 2011.
\newblock \doi{10.1016/j.physletb.2011.05.046}.

\bibitem[Deniz et~al.(2010)]{Deniz:2009mu}
M.~Deniz et~al.
\newblock {Measurement of Nu(e)-bar -Electron Scattering Cross-Section with a
  CsI(Tl) Scintillating Crystal Array at the Kuo-Sheng Nuclear Power Reactor}.
\newblock \emph{Phys. Rev.}, D81:\penalty0 072001, 2010.
\newblock \doi{10.1103/PhysRevD.81.072001}.

\bibitem[Mishra et~al.(1991)]{CCFR:1991lpl}
S.~R. Mishra et~al.
\newblock {Neutrino tridents and W Z interference}.
\newblock \emph{Phys. Rev. Lett.}, 66:\penalty0 3117--3120, 1991.
\newblock \doi{10.1103/PhysRevLett.66.3117}.

\bibitem[Vilain et~al.(1994)]{Vilain:1994qy}
P.~Vilain et~al.
\newblock {Precision measurement of electroweak parameters from the scattering
  of muon-neutrinos on electrons}.
\newblock \emph{Phys. Lett.}, B335:\penalty0 246--252, 1994.
\newblock \doi{10.1016/0370-2693(94)91421-4}.

\bibitem[Bellini et~al.(2011)]{Bellini:2011rx}
G.~Bellini et~al.
\newblock {Precision measurement of the 7Be solar neutrino interaction rate in
  Borexino}.
\newblock \emph{Phys. Rev. Lett.}, 107:\penalty0 141302, 2011.
\newblock \doi{10.1103/PhysRevLett.107.141302}.

\bibitem[Dror(2020)]{Dror:2020fbh}
Jeff~A. Dror.
\newblock {Discovering leptonic forces using nonconserved currents}.
\newblock \emph{Phys. Rev. D}, 101\penalty0 (9):\penalty0 095013, 2020.
\newblock \doi{10.1103/PhysRevD.101.095013}.

\bibitem[Aoyama et~al.(2020)]{Aoyama:2020ynm}
T.~Aoyama et~al.
\newblock {The anomalous magnetic moment of the muon in the Standard Model}.
\newblock \emph{Phys. Rept.}, 887:\penalty0 1--166, 2020.
\newblock \doi{10.1016/j.physrep.2020.07.006}.

\bibitem[Borsanyi et~al.(2021)]{Borsanyi:2020mff}
Sz. Borsanyi et~al.
\newblock {Leading hadronic contribution to the muon magnetic moment from
  lattice QCD}.
\newblock \emph{Nature}, 593\penalty0 (7857):\penalty0 51--55, 2021.
\newblock \doi{10.1038/s41586-021-03418-1}.

\bibitem[Lewin and Smith(1996)]{Lewin:1995rx}
J.~D. Lewin and P.~F. Smith.
\newblock {Review of mathematics, numerical factors, and corrections for dark
  matter experiments based on elastic nuclear recoil}.
\newblock \emph{Astropart. Phys.}, 6:\penalty0 87--112, 1996.
\newblock \doi{10.1016/S0927-6505(96)00047-3}.

\bibitem[Strigari(2009)]{Strigari:2009bq}
Louis~E. Strigari.
\newblock {Neutrino Coherent Scattering Rates at Direct Dark Matter Detectors}.
\newblock \emph{New J. Phys.}, 11:\penalty0 105011, 2009.
\newblock \doi{10.1088/1367-2630/11/10/105011}.

\bibitem[Billard et~al.(2014)Billard, Strigari, and
  Figueroa-Feliciano]{Billard:2013qya}
J.~Billard, L.~Strigari, and E.~Figueroa-Feliciano.
\newblock {Implication of neutrino backgrounds on the reach of next generation
  dark matter direct detection experiments}.
\newblock \emph{Phys. Rev.}, D89\penalty0 (2):\penalty0 023524, 2014.
\newblock \doi{10.1103/PhysRevD.89.023524}.

\bibitem[Drees and Zhao(2022)]{Drees:2021rsg}
Manuel Drees and Wenbin Zhao.
\newblock {U(1)L\ensuremath{\mu}\ensuremath{-}L\ensuremath{\tau} for light dark
  matter, g\ensuremath{\mu} \ensuremath{-} 2, the 511 keV excess and the
  Hubble tension}.
\newblock \emph{Phys. Lett. B}, 827:\penalty0 136948, 2022.
\newblock \doi{10.1016/j.physletb.2022.136948}.

\bibitem[Altmannshofer et~al.(2016)Altmannshofer, Gori, Profumo, and
  Queiroz]{Altmannshofer:2016jzy}
Wolfgang Altmannshofer, Stefania Gori, Stefano Profumo, and Farinaldo~S.
  Queiroz.
\newblock {Explaining dark matter and B decay anomalies with an $L_\mu -
  L_\tau$ model}.
\newblock \emph{JHEP}, 12:\penalty0 106, 2016.
\newblock \doi{10.1007/JHEP12(2016)106}.

\bibitem[Borah et~al.(2017)Borah, Sadhukhan, and Sahoo]{Borah:2017dqx}
Debasish Borah, Soumya Sadhukhan, and Shibananda Sahoo.
\newblock {Lepton Portal Limit of Inert Higgs Doublet Dark Matter with
  Radiative Neutrino Mass}.
\newblock \emph{Phys. Lett. B}, 771:\penalty0 624--632, 2017.
\newblock \doi{10.1016/j.physletb.2017.06.006}.

\bibitem[Abdelhameed et~al.(2019)]{CRESST:2019jnq}
A.~H. Abdelhameed et~al.
\newblock {First results from the CRESST-III low-mass dark matter program}.
\newblock \emph{Phys. Rev. D}, 100\penalty0 (10):\penalty0 102002, 2019.
\newblock \doi{10.1103/PhysRevD.100.102002}.

\bibitem[Agnes et~al.(2018)]{DarkSide:2018bpj}
P.~Agnes et~al.
\newblock {Low-Mass Dark Matter Search with the DarkSide-50 Experiment}.
\newblock \emph{Phys. Rev. Lett.}, 121\penalty0 (8):\penalty0 081307, 2018.
\newblock \doi{10.1103/PhysRevLett.121.081307}.

\bibitem[Aprile et~al.(2018)]{XENON:2018voc}
E.~Aprile et~al.
\newblock {Dark Matter Search Results from a One Ton-Year Exposure of XENON1T}.
\newblock \emph{Phys. Rev. Lett.}, 121\penalty0 (11):\penalty0 111302, 2018.
\newblock \doi{10.1103/PhysRevLett.121.111302}.

\bibitem[Aprile et~al.(2019)]{XENON:2019rxp}
E.~Aprile et~al.
\newblock {Constraining the spin-dependent WIMP-nucleon cross sections with
  XENON1T}.
\newblock \emph{Phys. Rev. Lett.}, 122\penalty0 (14):\penalty0 141301, 2019.
\newblock \doi{10.1103/PhysRevLett.122.141301}.

\bibitem[Agnese et~al.(2019)]{SuperCDMS:2018gro}
R.~Agnese et~al.
\newblock {Search for Low-Mass Dark Matter with CDMSlite Using a Profile
  Likelihood Fit}.
\newblock \emph{Phys. Rev. D}, 99\penalty0 (6):\penalty0 062001, 2019.
\newblock \doi{10.1103/PhysRevD.99.062001}.

\bibitem[Leane et~al.(2018)Leane, Slatyer, Beacom, and Ng]{Leane:2018kjk}
Rebecca~K. Leane, Tracy~R. Slatyer, John~F. Beacom, and Kenny C.~Y. Ng.
\newblock {GeV-scale thermal WIMPs: Not even slightly ruled out}.
\newblock \emph{Phys. Rev. D}, 98\penalty0 (2):\penalty0 023016, 2018.
\newblock \doi{10.1103/PhysRevD.98.023016}.

\bibitem[Ade et~al.(2016)]{Planck:2015fie}
P.~A.~R. Ade et~al.
\newblock {Planck 2015 results. XIII. Cosmological parameters}.
\newblock \emph{Astron. Astrophys.}, 594:\penalty0 A13, 2016.
\newblock \doi{10.1051/0004-6361/201525830}.

\bibitem[Profumo et~al.(2018)Profumo, Queiroz, Silk, and
  Siqueira]{Profumo:2017obk}
Stefano Profumo, Farinaldo~S. Queiroz, Joseph Silk, and Clarissa Siqueira.
\newblock {Searching for Secluded Dark Matter with H.E.S.S., Fermi-LAT, and
  Planck}.
\newblock \emph{JCAP}, 03:\penalty0 010, 2018.
\newblock \doi{10.1088/1475-7516/2018/03/010}.

\end{thebibliography}

\end{document}